\documentclass[3p,times,procedia]{elsarticle}

\usepackage{ecrc}

\volume{00}

\firstpage{1}

\journalname{Computer Communications}

\runauth{}

\jnltitlelogo{Computer Communications}

\usepackage{amssymb}
\usepackage{amsthm,array,booktabs,tabularx,caption}

\usepackage[figuresright]{rotating}

\usepackage[english]{babel}
\usepackage[T1]{fontenc}    
\usepackage[bookmarks=false,draft]{hyperref} 
\usepackage{amsfonts}       
\usepackage{makecell}
\usepackage[cmintegrals]{newtxmath}
\usepackage{algorithm}
\usepackage{algorithmic}
\usepackage{stfloats}
\usepackage{xcolor, soul}
\sethlcolor{cyan}
\usepackage{etoolbox}
\usepackage{url}
\usepackage{paralist}
\usepackage{enumitem}

\usepackage{subcaption}
\DeclareMathOperator*{\argmax}{arg\,max}
\theoremstyle{definition}
\newtheorem{thm}{Theorem}[section]
\newtheorem{lem}{Lemma}[section]
\newtheorem{cor}{Corollary}[thm]
\theoremstyle{remark}
\newtheorem*{rem}{Remark}

\theoremstyle{definition}
\newtheorem{defi}{Definition}[section]

\begin{document}

\begin{frontmatter}



\dochead{}

\title{Multi-Agent Reinforcement Learning for Long-Term Network Resource Allocation through Auction: a V2X Application}


\author[hwaddr]{Jing Tan}
\author[hwaddr]{Ramin Khalili}
\author[hpiaddr]{Holger Karl}
\author[hwaddr]{Artur Hecker}
\address[hwaddr]{Huawei Munich Research Center, Germany}
\address[hpiaddr]{Hasso Plattner Institute, University of Potsdam, Germany}

\begin{abstract}
We formulate offloading of computational tasks from a dynamic group of mobile agents (e.g., cars) as decentralized decision making among autonomous agents. We design an interaction mechanism that incentivizes such agents to align private and system goals by balancing between competition and cooperation. In the static case, the mechanism provably has Nash equilibria with optimal resource allocation. In a dynamic environment, this mechanism's requirement of complete information is impossible to achieve. For such environments, we propose a novel multi-agent online learning algorithm that learns with partial, delayed and noisy state information, thus greatly reducing information need. Our algorithm is also capable of learning from long-term and sparse reward signals with varying delay. Empirical results from the simulation of a V2X application confirm that through learning, agents with the learning algorithm significantly improve both system and individual performance, reducing up to 30\% of offloading failure rate, communication overhead and load variation, increasing computation resource utilization and fairness. Results also confirm the algorithm's good convergence and generalization property in different environments.\footnote{Preprint of paper accepted by Computer Communications July 2022.}
\end{abstract}

\end{frontmatter}
\section{Introduction}
\label{sec:intro}

Vehicular network (V2X) applications are characterized by huge number of users, dynamic nature, and diverse Quality of Service (QoS) requirements \cite{masmoudi2019survey}. They are also computation-intensive, e.g., self-driving applications such as semantic segmentation trains and infers from large neural networks \cite{hofmarcher2019visual}, motion planning solves non-convex optimization problems in real-time \cite{claussmann2019review,badue2020self}. These applications currently reside in the vehicle's onboard units (OBU) for short latency and low communication overhead. Even with companies such as NVidia developing OBUs with high computation power \cite{oh2019hardware}, post-production OBU upgrades for higher on-board computation power are typically not commercially viable; and irrespective of local OBU power, the ability to offload tasks to edge/cloud via multi-access edge computing (MEC) devices increases flexibility, protecting vehicles against IT obsolescence. Hence, offloading is a key technique for future V2X scenarios \cite{europe6g,you2021towards,5gaausecase1,5gaausecase2}. 

Currently, computation offloading decisions are strictly separated between user side and operating side \cite{mach2017mobile}. Vehicles act as users and decide what to offload to optimize an individual goal, e.g., latency \cite{baidya2020vehicular} or energy efficiency \cite{loukas2017computation}. Apart from expressing their preference through a predefined, static and universal QoS matrix \cite{masdari2021qos}, users cannot influence how their tasks are prioritized. The operating side centrally prioritizes tasks and allocates resources to optimize a system goal that is based on the QoS matrix; but this goal is not always the same as the users' goals, e.g., task number maximization \cite{choo2018optimal} or load balancing \cite{vondra2014qos}. 

This separation between system and user goals poses problems for both user and operating side, especially in the V2X context. V2X users have private goals \cite{shivshankar2014evolutionary}, are highly autonomous \cite{martinez2010assessing}, reluctant to share information or cooperate, and disobedient to a central planner \cite{feigenbaum2007distributed}. They want flexible task prioritization and influence resource allocation without sharing private information \cite{li2019learning}. On the operating side, an edge cloud computing architecture introduces signaling overhead and information delay in updating site utilization \cite{mach2017mobile}; coupled with growing user autonomy and service customization, traditional centralized optimization methods for resource allocation become challenging due to unavailability of real-time information and computational intractability.

We, hence, need an interaction mechanism between user and operating side based on incentives, not rules, and an algorithm that makes decentralized decisions with partial and delayed information in a dynamic environment. There are several challenges with such a mechanism. Users may game the system, resulting in potentially worse overall and individual outcomes \cite{oh2008few}---the first challenge \textbf{C1} is therefore how to incentivize user behavior such that users willingly align their private goals to the system goal while preserving their autonomy. The second challenge \textbf{C2} is finding an algorithm that efficiently learns from partial information with just enough incentive signals, keeping information sharing at a minimum.

There are different types of learning algorithms for decentralized decision-making \cite{bowling2002multiagent,weinberg2004best,chang2007no}. However, they face the challenge \textbf{C3} to trade off optimality and convergence while keeping computation and communication complexity tractable \cite{feigenbaum2007distributed}. Moreover, in the cases where decisions have long-term effects that are only apparent after a variable delay and where short-term rewards conflict with long-term goals, we need a learning algorithm that connects current action to rewards in the distant future. The challenge \textbf{C4} is to learn towards long-term goals with delayed and sparse reward signals.

We propose a decentralized decision-making mechanism based on \emph{second-price sealed-bid auction} that successfully addresses these challenges. 

\begin{itemize}
\item \textbf{C1}: A bidder has no knowledge of other bidders' bidding prices and it only receives bidding outcome and final price (i.e.\ payment) as feedback signal---this befits our requirement to limit information sharing. Our mechanism also utilizes the feedback signal to incentivize cooperative behavior and speed up learning. 
\item \textbf{C2}: For the dynamic case, we use a multi-agent reinforcement learning (MARL) algorithm, for its ability to learn with partial, noisy and delayed information, and a single reward signal.
\item \textbf{C3}: The RL algorithm learns the best-response strategy updated in a fictitious self play (FSP). FSP addresses strategic users' adaptiveness in a dynamic environment by evaluating state information incrementally and by keeping a weighted historical record \cite{heinrich2015fictitious}; it is easier to implement than other methods such as \cite{bowling2002multiagent}, especially with a large state and action space. 
\item \textbf{C4}: Furthermore, we use a curiosity learning model to encourage learning with sparse reward signals and a credit assignment model that attributes a delayed reward to historical action sequences.
\end{itemize}

Although we use the V2X context as an example, we emphasize that our method is not restricted to V2X applications---it can be applied to other applications facing similar challenges. 

Our empirical results show that over time, the best-response strategies stabilize and lead to significantly improved individual and overall outcomes. We compare active (learning-capable) and passive (learning-incapable) agents in both synthetic and realistic V2X setups. The synthetic setup shows the performance of the generic learning algorithm that is applicable in many distributed resource allocation scenarios: it successfully incentivizes distributed autonomous users to contribute to any existing centralized resource allocation solution by letting the users prioritize their own tasks. In the realistic setup, V2X-specific factors such as varying vehicle arrival rate and speed, distance to the MEC and communication delay, as well as tasks based on self-driving applications are considered. Our algorithm demonstrates capability to generalize to very different, previously unseen environments without the need for retraining. Each user in the network has its own, constant-size model, and all shared information for modeling is of constant size as well. The distributed nature means it is easily scalable to huge number of users without increased complexity, making it a potential add-on to any existing centralized solutions at the MEC.

To summarize, our main contributions are:
\begin{itemize}
\item We formulate computation offloading as a decision-making problem with decentralized incentive and execution. The strategic players are incentivized to align private and system goals by balancing between competition and cooperation.
\item We introduce MALFOY, a distributed algorithm that learns based on delayed and noisy environment information and a single, immediate reward signal. Our solution requires to share much less information. We show using extensive simulation that agents with MALFOY outperform agents without learning capabilities on overall resource utilization, offloading failure rate, load variation and communication overhead. 
\item In a realistic setup based on a concrete mobility model and V2X applications (i.e.\ self-driving), we further demonstrate MALFOY's flexibility to utilize long-term, sparse extrinsic reward signals with varying delay; it optimizes decision strategy over a long time period. MALFOY with long-term goals further reduces failure rate and shows better generalization properties.
\item We open-source our code \cite{dracosource2} to encourage reproduction and extension of our work.
\end{itemize}

Sec.\ref{sec:related} summarizes related work, Sec.\ref{sec:modelproblem} introduces the system model and formulates the problem, Sec.\ref{sec:solution} proposes our solution, Sec.\ref{sec:eval} presents empirical results, Sec.\ref{sec:conclusion} concludes the paper.

\section{Related Work}
\label{sec:related}

\subsection{Decentralized Decision-Making}

Centralized approaches such as \cite{kuo2018deploying, agarwal2018joint} for resource allocation and \cite{lyu2016multiuser, chen2018task,choo2018optimal} for offloading are suited to core-network and data-center applications where powerful central admission control and assignment (ACA) units can be set up, and data can be relatively easily obtained. They are not the focus of our study. 

Previous studies of decentralized systems address some of the issues in centralized approaches. Authors of \cite{blocher2020letting, stefan2021tnsm} propose a distributed runtime algorithm to optimize system goals but disregard user preferences. \cite{kumar2014bayesian, kumar2015coalition,chen2015efficient} only consider cooperative resource-sharing or offloading. Although some game-theoretic algorithms naturally deal with \textit{decentralized incentives}, they often require complete information of the game to \textit{centrally execute} the desired outcome. For example, \cite{cardellini2016game} assumes all user and node profiles are known \textit{a priori}, and \cite{guo2018mobile} assumes users share information---these assumptions may not be plausible in practice. In other approaches, the complexity of a decentralized system is reduced. \cite{chen2014decentralized} only considers channel interference in order to model the problem as a potential game and guarantee an equilibrium. \cite{shams2014energy} only considers discrete actions. \cite{li2019learning} learns with partial information, but it reduces complexity by assuming a single service type and constant arrival rate. \cite{khaledi2016optimal} also only considers discrete actions and single service type.

Classic decentralized decision-making mechanisms include dynamic pricing, negotiations, and auctions. Among these mechanisms, auction is most suitable in a dynamic and competitive environment, where the number of users and their preferences vary over time and distribution of private valuations is dispersed \cite{schindler2011pricing,einav2018auctions}. Auction is common in e.g.\ networking \cite{xu2012resource,xu2012interference}, energy \cite{lucas2013renewable}, and e-commerce \cite{huang2011design} for its efficient price discovery in a dynamic market with partial information. Among various forms of auction, second-price sealed-bid auction maximizes welfare rather than revenue and has limited information-sharing, hence befitting the requirements in our study. Specifically, our approach is based on Vickrey-Clarke-Groves (VCG) for second-price combinatorial auction \cite{vickrey1961counterspeculation}. Unlike \cite{jiang2015data} and \cite{li2021double}, which use VCG auction mechanism for resource allocation in a stationary environment, our agents react to other agents' behaviors and learn best response strategy in a dynamic environment. As in \cite{tan2022multi}, we use simultaneous combinatorial auctions as a simplified version of VCG---each bidder bids for all commodities separately, without having to specify its preference for any bundle \cite{feldman2013simultaneous}. Since it assumes no correlation between commodities, the simplification befits our study of independent service requests.

\subsection{Suitable Algorithms}

Among algorithms for decentralized decision-making, \emph{no-regret algorithms} apply to a wide range of problems and converge fast; however, they require to know best strategies that are typically assumed to be static \cite{chang2007no}. \emph{Best-response algorithms} search for best responses to other users' strategies, not for an equilibrium---they therefore adapt to a dynamic environment but they may not converge at all \cite{weinberg2004best}. To improve the convergence property of best-response algorithms, \cite{bowling2002multiagent} introduces an algorithm with varying learning rate depending on the reward; \cite{weinberg2004best} extends the work to non-stationary environments. But both these algorithms provably converge only with restricted classes of games. Besides these algorithms, independent learner methods~\cite{tan1993multi} such as also proposed for resource allocation \cite{cui2019multi} are used to reduce modeling and computation complexity, but they fail to guarantee equilibrium \cite{yang2018mean} and have overfitting problems \cite{lanctot2017unified}. Finally, federated learning \cite{mcmahan2017fl} is not applicable as it provides a logically centralized learning framework.

RL algorithms are well-known for their ability to learn sequential tasks and balance between exploitation and exploration \cite{teng2013reinforcement,almasri2020dynamic}. In our previous work \cite{tan2022multi}, we proposed a distributed RL algorithm to learn the best response strategy based on immediate reward signals, in a continuous state-action space. In \cite{tan2022multi}, RL is combined with supervised learning in a fictitious self-play (FSP) method to improve its convergence properties. Although it performed well, the algorithm ignores long-term effects of decision-making. 

This is because RL algorithms are typically ``short-term'' algorithms: \cite{minsky1961steps} first mentioned the necessity and difficulty of long-term temporal credit assignment in RL---it is essential to associate long-term reward to specific behavior or series of behaviors, such that behaviors that contribute to the long-term reward are prioritized. In RL algorithms with no focus on temporal credit assignment, importance of the immediate reward heavily outweighs estimated reward in the distant future, and the estimation has a bias that is related to the length of delay and exponential to the number of possible states \cite{arjona2018rudder}. Worse still, if the reward is both delayed and sparse, the reward estimation often has a high variance due to lack of predictable future states, especially with a big state-action space and high variance in the value of next states \cite{mataric1994reward,shahriari2017generic}. When decisions have long-term effects, such ``short-term'' algorithms would lead to worse performance. It proves to be one of the biggest challenges of applying RL in the real world \cite{dulac2021challenges}.
 
\subsection{Delayed and Sparse Rewards}
 
One common approach in long-term RL is to extract features from historical records, thus linking the delayed reward to behaviors in the past \cite{hester2013texplore}. Learning with such algorithms is inefficient since learning from past experiences can only happen when the delayed outcomes become available. To address the delay, \cite{mann2018learning} factorizes one state into an intermediate and a final state with independent transition probabilities and predicts each state at different intervals. \cite{hung2019optimizing} describes a credit-assignment method that focuses on the most relevant memory records via content-based attention; the algorithm is capable of locating past memory to execute new tasks and generalizes very well. These approaches focus more on the delay in reward signal and less on sparsity. In our setup, the long-term reward is delayed, sparse and sporadic, making these approaches inapplicable. 

To address sparsity of rewards, many model-based methods add intrinsic, intermediate rewards between sparse extrinsic reward signals. Such methods often adopt a supervised learning algorithm to predict next states and use the difference between the predicted and target state-action pair values as intrinsic reward. Although they propagate prediction inaccuracy into the future, they learn faster. For example, \cite{hester2013texplore} separately trains many ``feature models'' to predict each feature of the next state as well as a ``reward model'' to predict reward. Between sparse extrinsic rewards, the algorithm samples estimated next state and reward from the models. The models are only updated when there is new input available. Their approach assumes that state features are independent and can be learned separately, and the accuracy of the reward model is still related to the sparsity of the reward signal. \cite{pathak18largescale} uses a long-short-term memory (LSTM) to extract features from past memory that are more relevant to the current task, thus improving the model's generalization properties. The algorithm also uses two independent models to predict next state and action, the prediction accuracy becomes intermediate, intrinsic rewards inserted between sparse extrinsic rewards. In this approach, the intrinsic reward signal is not related to the extrinsic sparse reward and the final outcome of the game is not credited to specific agent behaviors. The lack of temporal credit assignment on a long time horizon affects learning efficiency \cite{minsky1961steps}, especially with sparse rewards and conflict between the agent's short-term and long-term goals \cite{khadka2018evolution,ijcai2020-368}, as is the case in our setup.

The credit assignment in \cite{khadka2018evolution} does not directly credit behaviors, but credit a population of models, therefore it requires each model to play a full episode in each step to generate experience. It is not applicable in our setup: a dynamic multi-agent environment with no clear episodes. \cite{ijcai2020-368} uses an attentional network to assign weights to past behaviors through reward-shaping. They focus on offline-learning of an independent credit assignment algorithm and decompose the long-term reward to densify reward signals; we need an online-learning algorithm in our dynamic environment, our learning agent learns more than just the credit assignment, and in our setup with conflicting short-term and long-term rewards, decomposed long-term rewards cannot be used directly to densify reward signals.

\section{System Model and Problem Formulation}
\label{sec:modelproblem}

\subsection{System model}
\label{sec:model}

\begin{figure*}[t]
	\centering
	\begin{minipage}{0.95\linewidth}
	\subcaptionbox{Example topology \label{topo}}{\includegraphics[width=0.24\linewidth]{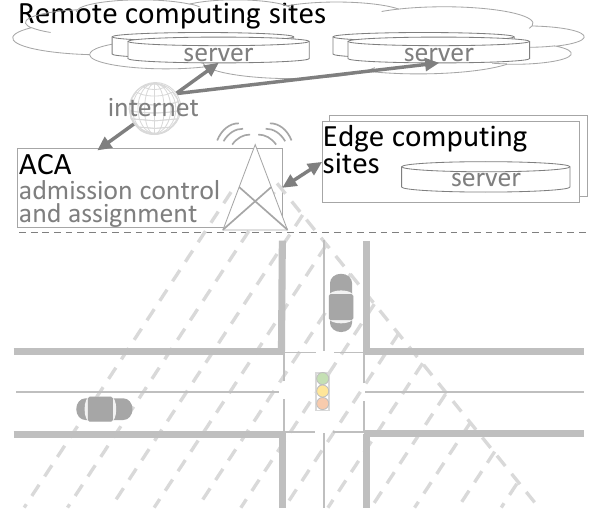}}
	\subcaptionbox{Message sequence \label{flow}}{\includegraphics[width=0.72\linewidth]{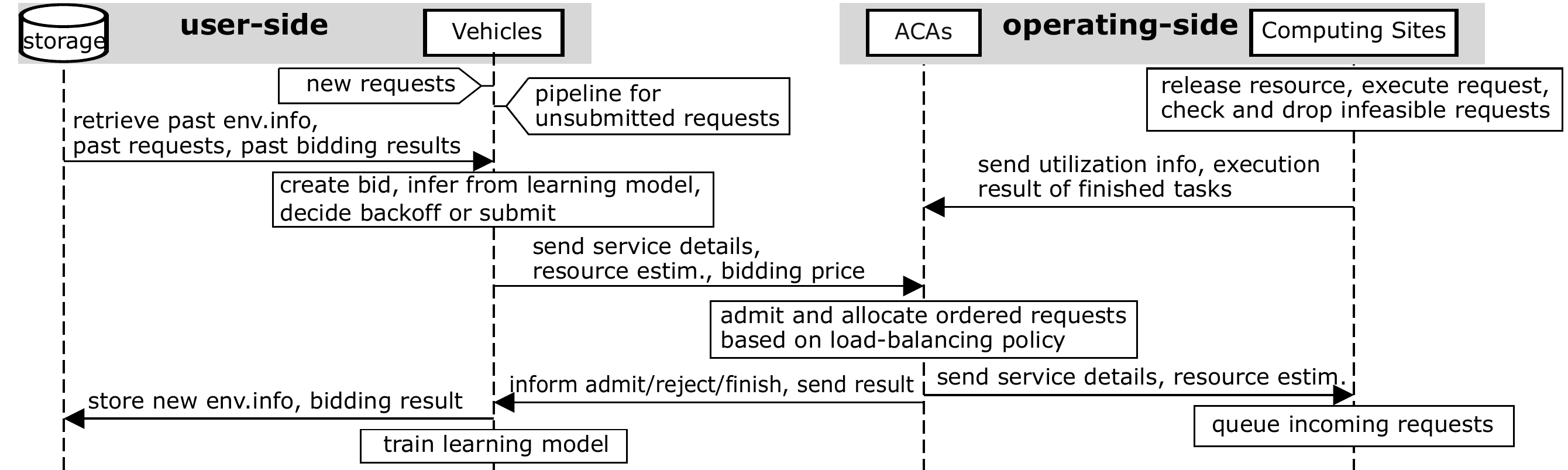}}
	\end{minipage}
	\vspace{-0.2cm}
	\caption{System model}
	\label{model}
\end{figure*}

Our system adopts the classic edge cloud computing architecture: user-side vehicles request services such as semantic segmentation and motion planning; operating-side ACAs (e.g., road-side units or base station) control admission of service requests and assign them to different computing sites, which own resources and execute services~\cite{whaiduzzaman2014survey} (Fig.\ref{topo}). 
We propose changes only to \begin{inparaenum}[1)] \item the algorithm admitting and assigning service requests and \item the interaction mechanism. \end{inparaenum} In addition, most signaling needs in our proposed approach are covered by the ISO 20078 standard on extended vehicle web services \cite{iso20078}; additional fields required to pass bidding and final price information are straightforward to implement. Channel security is not the focus of this study.

We first define a \emph{service request}; then, we explain in detail the user side and the operating side.

\subsubsection{Service request as bid}
\label{subsubsec:servicerequest}

The cloud-native paradigm decomposes services into tasks that can be deployed and scaled independently~\cite{alliance2019service}. A service request comprises \begin{inparaenum}[1)] \item a task chain, with varying number, type, order and resource needs of tasks, and \item a deadline. \end{inparaenum} We consider a system with custom-tailored services placed at different computing sites in the network; the properties of these services are initially unknown to the computing sites. This enables us to extend the use cases into new areas, e.g., self-driving~\cite{5gaausecase1,5gaausecase2}. We consider independent services, e.g., in self-driving, segmentation and motion planning can be requested independently. The corresponding class of the service is a \emph{service type}. 

Using motion planning as an example: every 100 milliseconds, the vehicle receives a service request of type ``motion planning'', which includes a task chain of two steps: localization and optimization. For execution, the vehicle uploads to the MEC the required input data (odometry, GPS, road image segments, etc.) that is estimated to be around 0.4Mbits. A high-definition map containing information of static object positions and labels can be stored on the access point and shared by all vehicles. After execution, the vehicle receives predicted optimal position and odometry for the next 3 seconds, estimated downlink size is ca. 6kbits \cite{broggi2014proud}. The vehicle expects this result to be sent back within 100 milliseconds (service deadline). 

We conceive of a vehicle's service request as a \emph{bid} in an auction. Besides the service request details, a bid includes the bidding price and the vehicle's estimated resource needs. 

\subsubsection{User side} 

Our study focuses on the behavior of the independent vehicles, conceived of as \emph{agents}. They act independently and do not share information. As a bidder, the vehicle bids for a \emph{commodity}---a service slot with necessary resources to execute the service request. The vehicle has a private \emph{valuation} for each commodity (i.e.\ the benefit it derives from winning the commodity), and its direct \emph{payoff} from the auction is its valuation minus the price it pays the seller for the commodity. Aside from the direct payoff, it also has other costs, and the total \emph{utility} is the sum of payoff and all costs. The vehicle's decision objective is to maximize average utility from joining the auction. If the vehicle bids a low price and loses, it suffers costs including transmission delay and communication overhead for bidding and rebidding; if it bids a high price and wins, it has reduced payoff. For a possibly lower cost or better payoff in the future, it can decide to join the auction at a later time (i.e.\ to back off). However, if backoff is too long, the vehicle has pressure to pay more and prioritize its request. Therefore, the vehicle balances between two options: i) back off and try later or ii) submit the bid immediately to the ACA unit for admission check. With the backoff option, \begin{inparaenum}[1)] \item vehicles are incentivized to balance between backoff and bidding through a cost factor; \item backoff time is learned, not randomly chosen; \item learning is based on vehicle state information (Sec.\ref{payment}).\end{inparaenum}

For example: when a vehicle receives a service request of the type ``motion planning'', it collects past bidding results and current environment parameters e.g., the number of other vehicles in the vicinity. Then it inputs service-specific information (deadline, estimated resource needs, input and output data size, etc.), historical data, and environment parameters to its onboard learning model, to infer the best bidding strategy for the motion planning request at the current time step. Transmission delay is calculated based on input and output data size. 

We study the learning algorithm in each vehicle. We use passive, non-learning vehicles as benchmark, to quantify the effect of learning on performance. Learning essentially sets the priority of a service request. This priority is used by the ACA to order requests; it is simply constant for non-learning agents, resulting in first-in, first-out processing order.

\subsubsection{Operating side} 

The ACA unit and computing sites are the operating side (Fig.\ref{flow}). The ACA unit decides to admit or reject ordered service requests. Upon admission, it assigns the request to a computing site according to a load-balancing policy. Due to information delay, execution uncertainty, system noise, etc., the resource utilization information at different sites is not immediately available to the ACA unit. If all computing sites are overloaded, service requests are rejected. For a rejected request, a vehicle can rebid a maximum number of times. If the request is admitted but cannot be executed before its deadline, the computing site drops the service and informs the ACA unit. Vehicles receive feedback on bidding and execution outcome, payment, and resource utilization (Sec.\ref{payment}).

The operating side does not have \emph{a priori} knowledge of the type, priority, or resource requirements of service requests. For example, once a site receives a previously unknown service, it uses an estimate of resource needs provided by the vehicle. Over time, a site updates this estimate from repeated executions of the same service. This enables a computing site to execute previously unseen service requests, based on a simple statistical estimation of resource needs. Extension to a more sophisticated form of learning is left to future work. 

The total service time of a request is the sum of processing, queueing, and transmission time. Each computing site may offer all services but with different resource profiles (i.e., amount and duration needed of CPU and memory), depending on the site's configuration. Site capacity is specified in abstract time-resource units: one such unit corresponds to serving one volume of request in one time unit at a server, when given one resource unit (in Sec. \ref{sec:eval}, we explain the detailed assumptions in simulation).

\subsection{Problem formulation}
\label{sec:problem}

\begin{table}[t]
 \centering
 \captionof{table}{Sec.\ref{sec:problem} and \ref{payment} symbol definition}
 \label{tab:problem}
 \begin{tabular}{c l c l c l c l c l c }
 Sym & Description & Sym & Description & Sym & Description\\
 \toprule
 $k \in K$ & service type/commodity & $n_k$ & $k$'s availability & $i \in I$ & service request/bid\\
 $m \in M$ & vehicle/bidder & $h \in H$ & resource types & $\omega_{i,h}$ & $i$'s requirement of $h$\\
 $B$ & wealth/budget & $v$ & bid value & $\beta$ & utilization\\
 $Q$ & service deadline & $\alpha$ & backoff decision & $b$ & bidding price\\
 $c$ & cost to join the auction & $q$ & backoff cost & $p$ & payment\\
 $z$ & bidding outcome & $u$ & immediate utility & $U$ & cumulated utility\\
 \bottomrule 
 \end{tabular}
\end{table} 

Table~\ref{tab:problem} summarizes the notation for this section. Let $M$ be the set of vehicles (bidders) and $K$ the set of commodities (service types), each type with total of $n_k^t$ available service slots at time $t$ in computing sites. Bidder $m$ has a reserve pool of wealth with an initial wealth of $B_m^0$. It has at most $1$ demand for each service type $k \in K$ at $t$, denoted by $m_k^t \in \{0,1\}$. It draws its actions for each service---whether to back off $\mathbf{\alpha}_m^t =\{\alpha^t_{m,1}, \cdots, \alpha^t_{m,|K|} \} \in \{0,1\}^{|K|}$, and which price to bid $\mathbf{b}_m^t =\{b^t_{m,1}, \cdots, b^t_{m,|K|} \} \in \mathbb R_+^{|K|}$---from a strategy. The bidding price is some unknown function $f_m$ of $m$'s private valuation of the service type $v_{m,k} \in \mathbb R_+$ and lower than or equal to the current amount $B^t$ in the reserve pool: $b_{m,k}^t=f_m(v_{m,k})$. The competing bidders draw their actions from a joint distribution $\pi_{-m}^t$ based on $(\mathbf{p}^1,\cdots, \mathbf{p}^{t-1})$, where $\mathbf{p}^t \in \mathbb R_+^{|K|}$ is the payment vector received at the end of time $t$, its element $p_k^t$ is the $(n_k^t+1)^{\textrm{th}}$ highest bid for service type $k$. If $m$ wins the bid for $k$, bidding outcome $z_m^t=1$, $m$ observes the new $\mathbf{p}^t$ as feedback, and receives an immediate utility $u_m^t$, which is a function of $m$'s private value $v_{m,k}^t$ of $k$, its bidding price $b_{m,k}^t$, and $p_k^t$; all losing bidders suffer $c_m^t$ as cost to join the auction, and bidders that backed off suffer $q_m^t$ as cost of backoff. We therefore write the immediate utility as $u_m^t=g(v_m^t,\mathbf{b}_m^t,z_m^t,\mathbf{p}^t,c_m^t,q_m^t)$. The auction repeats for $T$ periods. The goal is to maximize long-term cumulated utility: $U=\frac{1}{T} \sum\limits_{t=1}^T \sum\limits_{m \in M} u_m^t, T\to \infty$.

For any $k$, when availability $n_k^t<\sum\limits_{m \in M} m_k^t$, there is more demand than available service slots and we call it ``high contention''. When $n_k^t \geq \sum\limits_{m \in M} m_k^t$, we call it ``low contention''. In a dynamic environment, available service slot $n_k^t$ depends on utilization at $t-1$ and existing demand at $t$. Our setup imitates the noise and transmission delay in a realistic environment, which makes site utilization information outdated when it becomes available to the ACA unit for admission control (Fig.\ref{flow}). In Sec.\ref{sec:eval}, we demonstrate the algorithms' ability to learn despite outdated information.

Ideally, an auction is incentive-compatible. Unfortunately, with budget constraint and costs, the second-price auction considered here is no longer incentive-compatible. But we still use this type of auction as we have shown in our previous work \cite{tan2022multi} that it maximizes social welfare and optimally allocates resources. We also use the payment signal as additional feedback from ACA to aid bidders' learning process (Sec.\ref{modelDescription}).

\section{Proposed solution}
\label{sec:solution}

To solve the long-term reward maximization described in Sec.\ref{sec:problem}, we propose MALFOY: \textbf{M}ulti-\textbf{A}gent reinforcement \textbf{L}earning \textbf{FO}r sparse and dela\textbf{Y}ed reward; its ability to learn based on rewards with random delay makes it an extension to our previous work on short-term algorithms \cite{tan2022multi}. With this extension, the algorithm is generalized to target a wider range of problems, and the problem tackled in \cite{tan2022multi} becomes a special case where the long-term reward signals have an interval of $1$ (i.e.\ available at the end of every auction round).

In Sec.\ref{payment} we define a bidder's utility function and briefly explain the mechanism's theoretical properties in the static case. Then, we introduce MALFOY for the dynamic environment in Sec.\ref{fsp}-\ref{creditassign}. 

\subsection{Utility function}
\label{payment}

In this section, we first build up the utility function based on the payoff of classic second-price auction. Then, we add costs for backoff and losing the bid, incentivizing tradeoff between higher chance of success and lower communication overhead. Finally, we add the system resource utilization goal to the utility. 

In each auction round, if a bid $i$ for service type $k$ is admitted, its economic gain is $(v_{i,k}-p_{i,k})$. For each $k$, the bidder has a given private valuation $v_{i,k}$ that is \begin{inparaenum}[1)] \item linear to the bidder's estimated resource needs for the service request and \item within its initial wealth $B_m^0$. \end{inparaenum} The first condition guarantees Pareto optimality, the second condition avoids overbidding under rationality \cite{tan2022multi}. Our study does not consider irrational or malicious bidders, e.g., whose goal is to reduce social welfare even if individual outcome may be hurt. 

ACA records $b_{j,k}$ of the highest losing bid for each $k$, and sets the price to $p_{i,k}=b_{j,k}$. For $n_k$ available service slots, this would be the $n_k+1$th highest bidding price. For $n_k=1$, this would be the price of the second highest bid. Hence the name ``second-price auction''. If $i$ is admitted, the vehicle receives a payoff of $v_{i,k}-p_{i,k}$. If $i$ is rejected, it has a constant cost of $c_{i,k}$. The bidder's utility $\mathcal{u}_{i,k}$ so far:

\begin{flalign}\label{eq:uik}
\mathcal{u}_{i,k} =z_{i,k} \cdot (v_{i,k}-p_{i,k})-(1-z_{i,k}) \cdot c_{i,k} 
\end{flalign}

\noindent where $z_{i,k}=1$ means bidder wins bid $i$ for a service slot of service type $k$, which implies $b_{i,k}$ is among the highest $n_k$ bids for $k$. Ties are broken randomly.

We add $\alpha_{i,k} \in \{0,1\}$ for backoff decision: bidder submits the bid if $\alpha_{i,k}=1$, otherwise, it backs off with a cost $q_{i,k}$:

\begin{flalign}\label{eq:rewardbackoff}
u_{i,k} = \alpha_{i,k} \cdot (\mathcal{u}_{i,k} -\mathbf{1}|_{p_{i,k}=0} \cdot v_{i,k}) + (1-\alpha_{i,k}) \cdot q_{i,k} 
\end{flalign}

\noindent where $\mathbf{1}|_\text{conditions}=1$ if the conditions are true, otherwise $0$.

Especially in high contention, more rebidding causes communication overhead, but less rebidding reduces the chance of success. With $c_{i,k}$, the utility incentivizes less rebidding to reduce system-wide communication overhead (\textbf{C1}). Together with $q_{i,k}$, the bidder is incentivized to trade off between long backoff time and risky bidding. In our implementation (Sec.\ref{sec:eval}), $\alpha$ is continuous between $0$ and $1$ and linear in the backoff duration.

To further align bidder objectives with system overall objectives (\textbf{C1}), we include system resource utilization $\beta$ in the utility. This is to incentivize bidders to minimize system utilization. Hence, the complete utility definition is:

\begin{flalign}\label{eq:reward1}
u_i = \sum\limits_{k \in K} u_{i,k} + W \cdot (1-\beta) 
\end{flalign}

$W$ is a constant that weighs the utilization objective. In low contention, there is adequate resource to accept all bids, bidding price is less relevant, and backoff decision becomes more important.

To calculate Eq.\ref{eq:reward1}, the bidder needs only these feedback signals: bidding outcome $z_{i,k}$, payment $p_{i,k}$ and system utilization $\beta$, addressing \textbf{C2}.

In our previous work \cite{tan2022multi}, we provided a short version of our proof that \begin{inparaenum}[1)] \item the outcome of the game is an NE and a maximization of social welfare; and \item in high contention with resource capacity limit, the outcome is also an optimal resource allocation (i.e.\ Pareto optimality). \end{inparaenum} In this study, we provide the full proof in Appendix (Sec.\ref{appendix:SPAwithpenalty} and \ref{appendix:paretoOptimal}). 

In a dynamic environment, MALFOY learns to achieve reward maximization using the utility function. The algorithm consists of three parts: \begin{inparaenum}[1)] \item the fictitious self-play (FSP) (Sec.\ref{fsp}), including an RL (Sec.\ref{modelDescription}) and a supervised learning (SL) model; \item the curiosity learning model (Sec.\ref{subsec:longterm}); and \item the credit assignment (Sec.\ref{creditassign}). \end{inparaenum} RL seeks to learn the best-response strategy in a huge state-action space by balancing between exploitation and exploration. To improve its convergence properties, a FSP is wrapped around the RL to stabilize the learning process. To further improve the model's generalization properties and learning efficiency with sparse extrinsic reward signal, we add a curiosity learning model to the FSP. Finally, to enhance the model's ability to learn from long-term, delayed extrinsic rewards, we add a credit assignment model to the FSP that attributes the long-term reward to short-term actions.

In our previous work \cite{tan2022learning}, we simulated two common repeated auctions with single commodity and let three types of algorithms compete directly against each other: the short-term FSP algorithm, the long-term FSP with curiosity learning, and the long-term FSP with both curiosity learning and credit assignment (same as MALFOY). Our results showed that MALFOY outperformed all others.

\begin{figure}[t]
		\centering
		\includegraphics[width=0.9\linewidth]{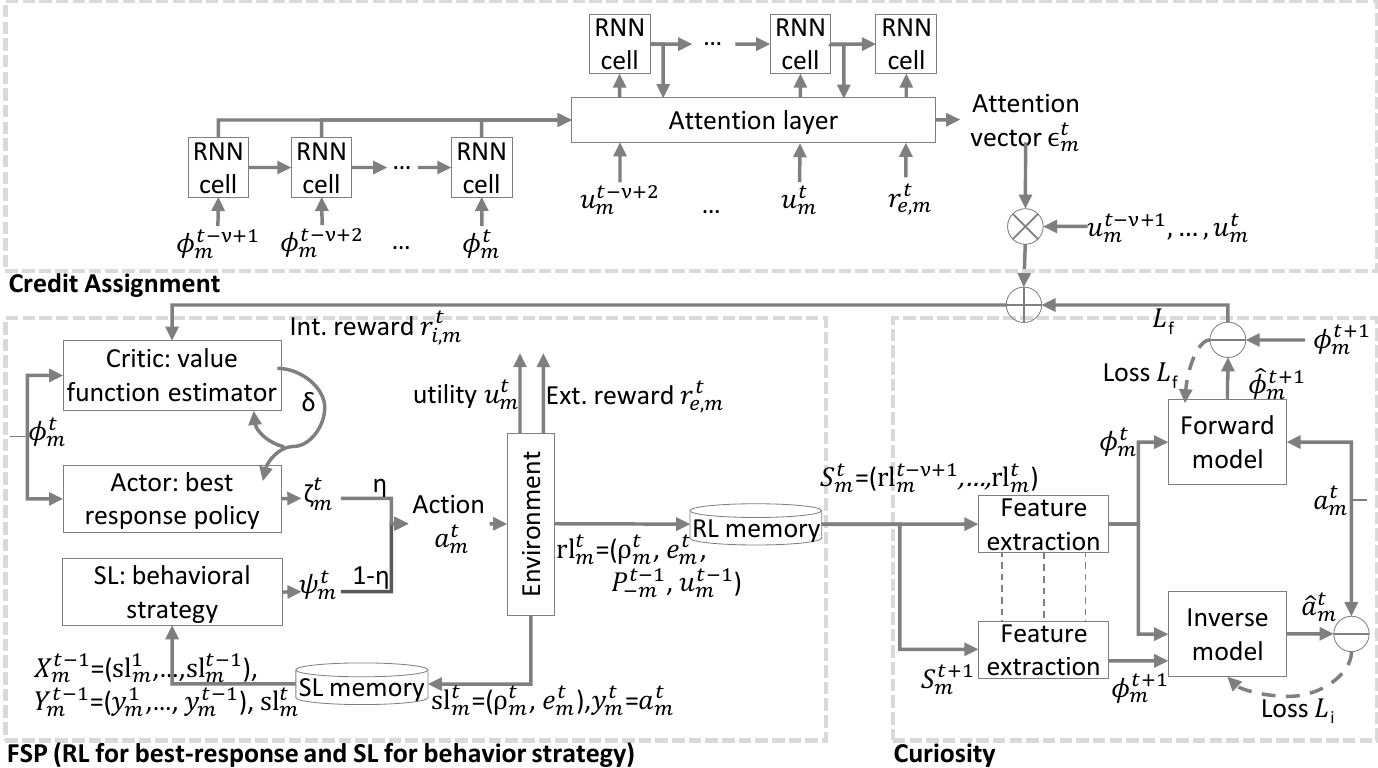}
		\caption{Long-term algorithms running at each vehicle}
		\label{attentionchart}
	\end{figure}

In the following sections, we explain the parts shown in Fig.\ref{attentionchart} in detail. Table \ref{tab:fsp} summarizes the notation.

\subsection{The FSP method}
\label{fsp}

	\begin{table}[t]
	 \centering
	 \captionof{table}{Sec.\ref{fsp}-\ref{creditassign} symbol definition}
	 \label{tab:fsp}
	 \begin{tabular}{c l c l c l c | c | c}
	 Sym & Description & Sym & Description & Sym & Description\\
	 \toprule
	 $\zeta$ & best response & $\psi$ & behavioral strategy & $\mathbf e_m$ & env. variables\\
	 $\rho$ & private bidder info & $\mathbf{a}$ & action, $\mathbf{a}=(\alpha, b)$ & $P_{-m}^t$ & other bidders state\\
	 $\text{sl}_m^t$ & SL present state & $\text{rl}_m^t$ & RL present state & $S_m^t$ & RL complete state\\
	 $\lambda$ & $\bar{u}$'s weight factor & $\theta$ & actor parameters & $\mathbf w$ & critic parameters\\
	 $\gamma$ & learning rate & $\delta$ & TD error & $\eta$ & $\zeta$'s weight\\
	 $\nu$ & history length & $\mu$ & action mean & $\Sigma$ & action covariance\\
	 $\phi$ & featurized state & $\epsilon$ & credit assign. weight & $r_{i}$ & intrinsic reward\\
	 $r_{e}$ & extrinsic reward & $L_{f}$ & forward mdl loss & $L_{i}$ & inverse mdl loss\\
	 $\xi$ & reward weight\\
	 \bottomrule 
	 \end{tabular}
	\end{table}
	\hfill

The fictitious self-play (FSP) method addresses the convergence challenge of a best-response algorithm (\textbf{C3}). FSP balances exploration and exploitation by replaying its own past actions to learn an average behavioral strategy regardless of other bidders' strategies; then, it cautiously plays the behavioral strategy mixed with best response \cite{heinrich2015fictitious}. The method consists of two parts: \begin{inparaenum}[1)] \item a supervised learning (SL) algorithm predicts the bidder's own behavioral strategy $\psi$, and \item an RL algorithm predicts its best response $\zeta$ to other bidders. \end{inparaenum} The bidder has $\eta,\lim\limits_{t \to \infty} \eta =0$ probability of choosing action $\mathbf{a}=\zeta$, otherwise it chooses $\mathbf{a}=\psi$. The action includes backoff decision $\alpha$ and bidding price $b$. If $\alpha$ is above a threshold, the bidder submits the bid; otherwise, the bidder backs off for a duration linear in $\alpha$. We predefine the threshold to influence bidder behavior: with a higher threshold, the algorithm becomes more conservative and tends to back off more service requests. Learning the threshold (e.g., through meta-learning algorithms) is left to future work.

Although FSP only converges in certain classes of games \cite{LESLIE2006285} (and in our case of a multi-player, general-sum game with infinite strategies, it does not necessarily converge to an NE), it is still an important experiment as our application belongs to a very general class of games; and empirical results show that by applying FSP, overall performance is greatly improved compared to using only RL. The FSP is described in Alg.\ref{algorithm}. 

Input to SL includes bidder $m$'s service requests---service type, resource amount required, and deadline: $\rho_m^t=\{(k_i, \omega_{i,h},Q_{i})| i \in I, h \in H\}$ ($m$ can create multiple bids, each an independent request for service type $k_i$; $\rho_m^t$ is the set of all $m$'s bids at $t$), current environment information visible to $m$, denoted $e_m^{t}$ (e.g., number of bidders in the network and system utilization $\beta^t$), and other bidder conditions, e.g. initial wealth $B^0$, and current wealth $B^t$. SL infers behavioral strategy $\psi_m^t$. The input $\text{sl}_m^t=(\rho_m^t,e_m^t)$ and actual action $\mathbf{a}_m^t$ are stored in SL memory to train the regression model. we use a multilayer perceptron in our implementation. 

Input to RL: $\phi_m^t$ is a featurized state vector from the original state vector $S_m^t$. The feature extraction module is part of curiosity model (Sec.\ref{subsec:longterm}); it extracts features that are most relevant to the agent's actions. In our algorithm, $S_m^t$ is constructed from $m$'s present state $\text{rl}_m^t$. $\text{rl}_m^t$ includes \begin{inparaenum}[1)] \item $\rho_m^t$; \item $e_m^t$; \item previous other bidders' state $P_{-m}^{t-1}$, represented by the final price $p_k$, or $P_{-m}^t=\mathbf{p}^t=\{ p_k^t| k \in K \}$; and \item calculated utility $u_m^{t-1}$ according to Eq.\ref{eq:reward1}. \end{inparaenum} To consider historical records, we take $\nu$ most recent states to form the complete state vector: $S_m^t=\{\text{rl}_m^\tau|\tau=t-\nu+1,\cdots,t\}$. Thus, input data consists mostly of information private to the user $m$, and the environment data, as well as past prices, are easily obtainable public information (\textbf{C2}). RL outputs best response $\zeta_m$. We provide a detailed description of the RL algorithm below.

\begin{figure}[t]
\begin{minipage}[t]{0.48\linewidth}
	  \begin{algorithm}[H]
	  \small
	  \begin{algorithmic}[1]
	  \STATE Initialize $\psi_m,\zeta_m$ arbitrarily,$\nu,t=1,\eta=1/t,P_{-m}^{t-1}=\mathbf{0}, u_m^{t-\nu+1},\cdots,u_m^{t-1}=0$, observe $e_m^t$, create $\text{rl}_m^t,\text{sl}_m^t$ and add to memory
	  \WHILE{true}
	    \STATE Take action $\mathbf{a}_m^t=(1-\eta)\psi_m^t+\eta \zeta_m^t$
	    \STATE Receive $P_{-m}^t$, calculate $u_m^t$, observe $\rho_m^{t+1},\mathbf e_m^{t+1}$
	    \STATE Create and add state to RL memory: $\text{rl}_m^{t+1}$
	    \STATE Create and add state to SL memory: $(\text{sl}_m^{t+1},\mathbf{a}_m^t)$
	    \STATE Construct $S_m^t,S_m^{t+1}$
	    \STATE Get $\phi_m^t,\phi_m^{t+1},r_{i,m}^t=\text{Curiosity}(S_m^t,S_m^{t+1},\mathbf{a}_m^t)$
	    \STATE Get $\zeta_m^{t+1}=\text{RL}(\phi_m^t,\phi_m^{t+1},r_{i,m}^t)$
	    \STATE Get $\psi_m^{t+1}=\text{SL}(\text{sl}_m^{t+1})$
	    \STATE $t \gets t+1$, $\eta \gets 1/t,\zeta_m^{t} \gets \zeta_m^{t+1},\psi_m^{t} \gets \psi_m^{t+1}$
	  \ENDWHILE
	  \end{algorithmic}
	  \caption{FSP algorithm for bidder $m$}
	  \label{algorithm}
	  \end{algorithm}
\end{minipage}
\hfill
\begin{minipage}[t]{0.48\linewidth}
	  \begin{algorithm}[H]
	  \small
	  \begin{algorithmic}[1]
	  \STATE Initialize $\theta, w$ arbitrarily. Initialize $\lambda$
	  \WHILE{true}
	    \STATE Input $t$ and $\phi_m^t,\phi_m^{t+1}$
	    \STATE Run critic and get $\hat V(\phi_m^{t}, \mathbf w),\hat V(\phi_m^{t+1},\mathbf w)$
	    \STATE Calculate $\bar r_{i,m}=\lambda \bar r_{i,m}$ and $\delta$
	    \STATE Run actor and get $\mu(\theta), \Sigma(\theta)$
	    \STATE Sample $\zeta_m^{t+1}$ from $F(\mu,\Sigma)$, update $\mathbf w$ and $\theta$
	  \ENDWHILE
	  \end{algorithmic}
	  \caption{RL algorithm for bidder $m$}
	  \label{algorithmRL}
	  \end{algorithm}
\end{minipage}
\end{figure}

\subsection{The RL Algorithm}
\label{modelDescription}

Authors of \cite{khaledi2016optimal} use VCG and a learning algorithm for the bidders to adjust their bidding price based on budget and observation of other bidders. Our approach is similar in that we estimate other bidders' state $P_{-m}$ from payment information and use the estimate as basis for a policy. Also, similar to their work, payment information is only from the seller.

Our approach differs from \cite{khaledi2016optimal} in several major points. We use a continuous space for bidder states (i.e., continuous value for payments). As also mentioned in \cite{khaledi2016optimal}, a finer-grained state space yields better learning results. Moreover, we consider multiple commodities, which is more realistic, and therefore has a wider range of applications. Further, we do not explicitly learn the transition probability of bidder states. Instead, we use historical states as input and directly determine the bidder's next action.

We use the actor-critic algorithm \cite{sutton2018reinforcement} for RL (Alg.\ref{algorithmRL}). The \textbf{critic} learns a state-value function $V(\phi)$. Parameters of the function are learned through a neural network that updates with $\mathbf w \gets \mathbf w + \gamma^w\delta \nabla \hat V(\phi, \mathbf w)$, where $\gamma$ is the learning rate and $\delta$ is the temporal difference (TD) error. For a continuing task with no terminal state, no discount is directly used to calculate $\delta$. Instead, the average reward is used \cite{sutton2018reinforcement}: $\delta =r-\bar r+\hat V(\phi',\mathbf w) - \hat V(\phi,\mathbf w)$. In our case, the reward is intrinsic reward $r_{i,m}$, which is utility $u_m$ weighted by their importance to the delayed extrinsic reward through weight vector $\epsilon$ from the credit assignment model (Sec.\ref{creditassign}). We use exponential moving average (with rate $\lambda$) of past rewards as $\bar r$.

The \textbf{actor} learns the parameters of the policy $\pi$ in a multidimensional and continuous action space. Correlated backoff and bidding price policies are assumed to be normally distributed: $F(\mu,\Sigma) = \frac{1}{\sqrt{|\Sigma|}} \exp(-\frac{1}{2}(\mathbf x-\mu)^T\Sigma^{-1}(\mathbf x - \mu))$. For faster calculation, instead of covariance $\Sigma$, we estimate lower triangular matrix $L$ ($LL^T=\Sigma$). Specifically, the actor model outputs the mean vector $\mu$ and the elements of $L$. Actor's final output $\mathbf{\zeta}$ is sampled from $F$ through: $\mathbf{\zeta} = \mu + L\mathbf{y}$, where $\mathbf{y}$ is an independent random variable from standard normal distribution. Update function is $\theta \gets \theta + \gamma^\theta \delta \nabla \ln \pi(\mathbf{a}|S,\theta)$. We use $\frac{\partial \ln F}{\partial \mu} =\Sigma (\mathbf x-\mu)$ and $\frac{\partial \ln F}{\partial \Sigma} = \frac{1}{2} (\Sigma(\mathbf x-\mu)(\mathbf x-\mu)^T\Sigma-\Sigma)$ for back-propagation.

The RL's objective is to find a strategy that, given input $\phi_m^t$, determines $\mathbf{a}$ to maximize $\frac{1}{T-t}\mathbb{E}[\sum_{t'=t}^T r_{i,m}^{t'}]$. To implement the actor-critic RL, we use a stacked convolutional neural network (CNN) with highway \cite{srivastava2015training} structure similar to the discriminator in \cite{yu2017seqgan} for both actor and critic models. The stacked-CNN has diverse filter widths to cover different lengths of history and extract features, and it is easily parallelizable, compared to other sequential networks. Since state information is temporally correlated, such a sequential network extracts features better than multilayer perceptrons. The highway structure directs information flow by learning the weights of direct input and performing non-linear transform of the input.

In low contention, authors of \cite{perkins2014game} prove that an actor-critic \cite{sutton2018reinforcement} RL algorithm converges to Nash equilibrium (NE) in a potential game. In high contention, although we prove the existence of an NE in the static case, the convergence property of our algorithm in a stochastic game is not explicitly analyzed. We show it through empirical results in Sec.\ref{sec:eval}.

Next, we describe the curiosity learning and credit assignment models in detail, which are key to the long-term algorithm.

\subsection{The Curiosity Model}
\label{subsec:longterm}

\begin{figure}[t]
\begin{minipage}[t]{0.48\linewidth}
	  \begin{algorithm}[H]
	  \small
	  \begin{algorithmic}[1]
	  \STATE Initialize model parameters, $\epsilon$ arbitrarily. Initialize $\xi$
	  \WHILE{true}
	    \STATE Input $a_m^t$ and $S_m^t,S_m^{t+1}$ constructed from RL memory
	    \STATE Run feature extraction, get $\phi_m^t$ and $\phi_m^{t+1}$
	    \STATE Run forward model, get $\hat \phi_m^{t+1}$, calculate $L_f$
	    \STATE Run inverse model, get $\hat a_m^t$, calculate $L_i$
	    \STATE Update model parameters
	    \STATE Infer from credit assignment, extract $\epsilon_m^t$ from attention layer
	    \STATE Calculate and output $r_{i,m}^t$
	  \ENDWHILE
	  \end{algorithmic}
	  \caption{Curiosity learning algorithm}
	  \label{algorithmcuriosity}
	  \end{algorithm}
\end{minipage}
\hfill
\begin{minipage}[t]{0.48\linewidth}	  
	  \begin{algorithm}[H]
	  \small
	  \begin{algorithmic}[1]
	  \STATE Initialize model parameters arbitrarily, initialize batch size $\nu$
        \STATE Input $r_{e,m}^t$ and $S_m^t,\cdots,S_m^{t-\nu+1}$, $u_m^t,\cdots,u_m^{t-\nu+2}$ from RL memory
	  \STATE Run feature extraction and get $\phi_m^t,\cdots,\phi_m^{t-\nu+1}$
	  \FOR{$\tau \gets t-\nu+1$ to $t-1$}
	    \STATE Input $\phi_m^\tau$ to encoder, get encoder output $\text{enc}_o$
	    \STATE Input $\text{enc}_o,u_m^{\tau+1}$ to decoder, get output $\text{dec}_o^\tau$
	  \ENDFOR
	  \STATE Input $\phi_m^t$ to encoder, get $\text{enc}_o$
	  \STATE Input $\text{enc}_o,r_{e,m}^t$ to decoder, get $\text{dec}_o^t$	
	  \STATE Update model params, output $\epsilon_m^t$ from attention layer
	  \end{algorithmic}
	  \caption{Credit assignment algorithm}
	  \label{algorithmattention}
	  \end{algorithm}
\end{minipage}
\end{figure}

Our curiosity model is based on the vanilla model from \cite{pathak2017curiosity}. They use feature extraction to identify features that can be influenced by the agent's actions, thus improving the model's generalization properties in new environments. In our competitive and dynamic environment, next state depends not only on the current state, but on a number of historical states. We therefore extract features from current and historical records $S_m^t$. The resulting featurized state vector $\phi_m^t=\text{feature}(S_m^t)$ is also the input to the RL (Sec.\ref{modelDescription}) and the credit assignment (Sec.\ref{creditassign}). 

\cite{pathak2017curiosity} uses a forward model and an inverse model to predict next state and next action, respectively. These are supervised learning models with the objective to minimize loss $L_f=\| \phi_m^t-\hat\phi_m^t \|_2^2$ and $L_i=\| \mathbf{a}_m^t-\hat{\mathbf{a}}_m^t \|_2^2$. One of the objectives of the forward and inverse models is to improve prediction accuracy of the consequence of the agent's actions, even without any reward signal. In our game setup, we have short-term intrinsic reward signals (only not aligned and potentially conflicting with the extrinsic rewards); therefore, we adapt the input to include the previous intrinsic reward values, and the forward model's objective is to improve prediction accuracy of both the state and the intrinsic reward. 

In \cite{pathak2017curiosity}, the intrinsic reward is the weighted loss of the forward model: $r_{i,m}^t=\xi L_f$, and the bigger the forward loss, the higher the intrinsic reward. Through the adversarial design, the model is encouraged to explore state-actions where the agent has less experience and prediction accuracy is low. The intrinsic rewards are inserted between sparse extrinsic rewards to improve learning efficiency despite the sparseness---the authors of \cite{pathak2017curiosity} call this internal motivation ``curiosity-driven exploration''. In our approach, we apply the same method with a modified intrinsic reward definition: $r_{i,m}^t=\xi L_f^t + (1-\xi) \epsilon u_{m}^t$, where $\xi$ is a predefined weight factor to balance between the two short-term objectives, and $\epsilon$ is a weight factor from the credit assignment model (see below). The objective is to maximize: $\mathbb{E}_\pi [\sum_t r_{i,m}^t]-L_i-L_f$. Pseudo code is in Alg.~\ref{algorithmcuriosity}. 

\subsection{Credit Assignment Model}
\label{creditassign} 

The credit assignment model uses a sequential network (recurrent neural network as encoder and decoder) with an attention layer. Typically, such a sequential network is used to identify correlation between sequenced input elements $\text{enc}_i$ and predict a corresponding sequence of output elements $\hat{\text{dec}}_o$. The sequential network is enhanced with an attention layer, which establishes relationship between any elements in the sequence, regardless of the distance between them. Our credit assignment model is inspired by \cite{ijcai2020-368}, our model is different in that we do not decompose the extrinsic reward. 

In our credit assignment model, we are not interested in predicting $\hat{\text{dec}}_o$. Instead, we want to determine the contribution of each state-action pair towards the final extrinsic reward $r_{e,m}^t$. Therefore, we trigger the training of the credit assignment model only when there is a new signal $r_{e,m}^t$ at time $t$: this signal becomes the last element of the target vector. We train the model on the batch of $\nu$ featurized state vectors $\text{enc}_i=\{\phi_m^{t-\nu+1},\cdots,\phi_m^{t}\}$ with both short- and long-term rewards as target vector, $\text{dec}_o = \{u_{m}^{t-\nu+2},\cdots,u_{m}^t,r_{e,m}^t\}$. In time step $\tau \in [t-\nu+1,t]$, the attention layer generates a weight vector corresponding to input vector $\text{enc}_i$, marking its relevance to the current output prediction $\hat{\text{dec}}_o^\tau$, until in the last time step $t$, the attention layer outputs a weight vector $\epsilon_m^t=\{\epsilon_1,\cdots,\epsilon_\nu|\sum_{i=1}^n \epsilon_i=1\}$ corresponding to $\text{enc}_i$ that marks their relevance to the last output $r_{e,m}^t$. Model parameters are updated with the mean square error between the generated output $\hat{\text{dec}}_o$ and target vector $\text{dec}_o$.

The weight vector $\epsilon_m^t$ is then multiplied with the original utilities $u_m^t$. Through $\epsilon_m^t$, short- and long-term rewards are aligned, even if they are conflicting in nature. Between sparse extrinsic rewards, only the forward network of credit assignment model is run to infer a weight vector.

The features that make our algorithm truly long-term are: \begin{inparaenum}[1)] \item reward prediction, \item more exploration in the early stages of learning, and \item short- and long-term reward alignment through credit assignment. \end{inparaenum} Points 1) and 2) are achieved through an adapted curiosity model (Sec.\ref{subsec:longterm}). Point 3) is achieved through a hierarchical structure that uses an attentional network to learn and assign weights to short-term rewards based on their relevance to the long-term, sparse extrinsic reward; the learning process is only triggered when a new extrinsic reward becomes available (Sec.\ref{creditassign}). Between the extrinsic reward signals, the FSP+curiosity model learns to better predict next states, actions, and intrinsic rewards (\textbf{C4}). 

In our setup, only the extrinsic reward is delayed; for the intrinsic reward, we measure offloading failure at the time of task admission. However, our algorithm can also learn with delayed intrinsic rewards, e.g., if the measurement of offloading failure is after task execution. For the sake of simplicity, we assume that failure rate measured before and after actual task execution is the same. We verify this assumption in the next section, where we show that applying our solution, we reach a system responsiveness \cite{avizienis2004basic} of 99\% (i.e. 99\% of the admitted jobs at MEC are successfully processed before their deadlines).

\section{Evaluation}
\label{sec:eval}

We develop a Python discrete-event simulator, with varying number of vehicles of infinite lifespan, one MEC with ACA and edge computing site, and one remote computing site (extension to multiple ACA units and computing sites is left to future work). The edge and remote sites have different resource profiles. To imitate a realistic, noisy environment, the remote site is some distance to the ACA unit, such that data transmission would cause non-negligible delay in state information update. We also add a noise to the delay and to the actual resource need that is independently drawn from a normal distribution (we define the parameters of the normal distributions before the simulation. The analysis of the impact of different simulation parameters is left to future work). Each vehicle is randomly and independently initialized with a budget of ``high'' or ``low'' with 50\% probability. For the operating-side load-balancing policy, we apply state-of-the art resource-intensity-aware load-balancing (RIAL) \cite{8006307} with slight modifications. The method achieves dynamic load-balancing among computing sites through resource pricing that is correlated to the site's load, and loads are shifted to ``cheaper'' sites. The queueing time and processing time of each service request is initialized with a constant, and as the computing site processes more of the same service type, the estimated queueing and processing time is drawn from the empirical distribution of past observations. Finally, we compare the performance of active agents (MALFOY on the user side, RIAL on the operating side, M+R) to passive agents (only RIAL on the operating side), as shown in Fig.\ref{flow}. Evaluation data is collected from additional evaluation runs after the models are trained, with random incoming service requests newly generated by a two-state Markov-modulated Poisson process (MMPP) \cite{wang2013characterizing}. 

We evaluate the following metrics:
\begin{itemize}
\item \textbf{Offloading failure rate (OFR):} Ratio of offloading requests rejected by ACA during admission control. As mentioned in Sec. \ref{creditassign}, we observe M+R's responsiveness of $99\%$, which is consistently higher than RIAL for all results in the paper. Therefore, this is a close approximation of the ratio of failed offloading requests that are either rejected, or not executed within deadline.
\item \textbf{Resource utilization:} Ratio of resources effectively utilized at computing sites = (sum of utilized resource units in  all resource types and all computing sites in the current time step) / (sum of total resource units in all resource types and all computing sites at any time). 
\item \textbf{Rebidding overhead:} If a bid is rejected before deadline, the vehicle can bid again. More rebidding causes communication overhead, but less rebidding reduces the chance of success. We study this tradeoff, comparing the average number of actual rebiddings per vehicle within maximum permitted-rebidding (MP).
\end{itemize}

\begin{figure*}[t]
	\centering
	\subcaptionbox{Failure rate comparison with MP of 1 and 5 times, respectively. The x-axis is system resource capacity: to the right, system capacity increases, creating a low-contention scenario. M+R\_1 reduces OFR by $40\%$, achieves $1\%$ OFR in low contention; RIAL only reaches $2\%$.\label{successbycapa}}{\includegraphics[width=0.45\linewidth]{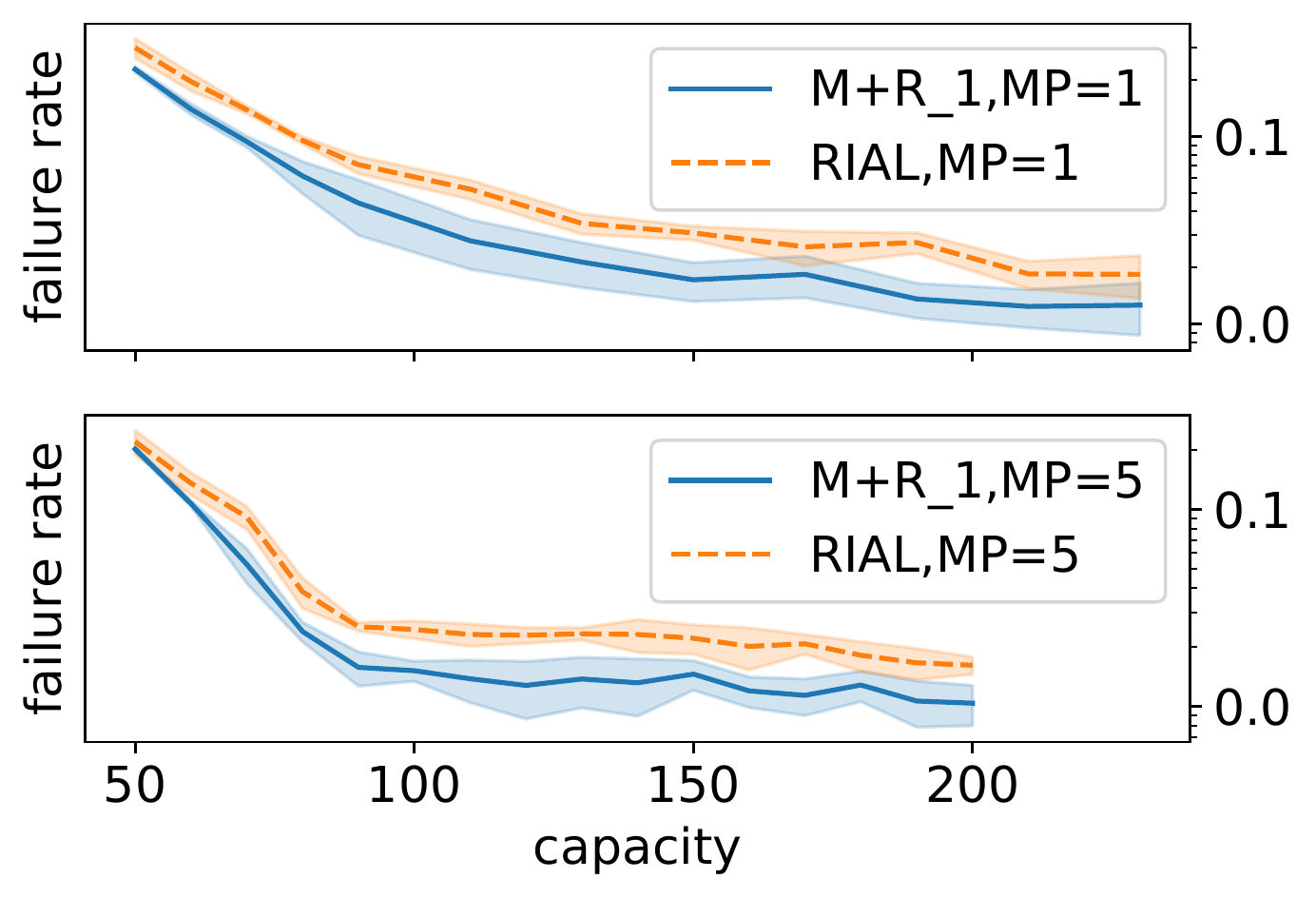}}\hspace{1em}
	\subcaptionbox{Comparison of resource capacity needs, when given OFR service level requirements. X-axis is required OFR level: to the right, stricter requirement of low failure rate applies, and resource capacity is increased to meet requirement. For the same OFR, M+R\_1 needs much less resource, e.g., for $2\%$ OFR and MP=1, M+R\_1 needs $38\%$ less resource.\label{capause}}{\includegraphics[width=0.45\linewidth]{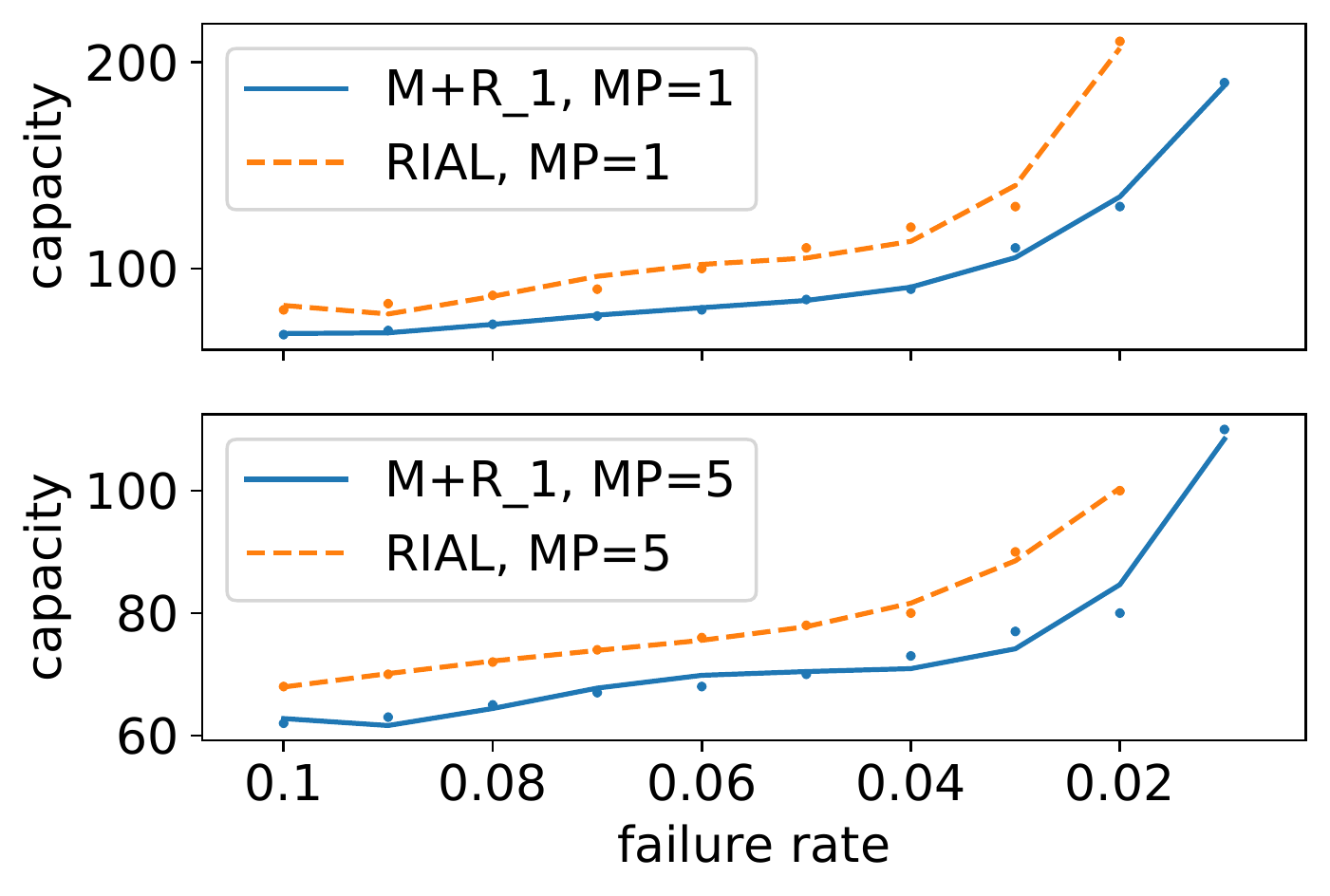}}\hfill
	\subcaptionbox{Rebidding overhead comparison with MP=1. X-axis is resource capacity, to the right is low contention with high capacity. M+R\_1 reduces rebidding overhead by $32\%$ on average. \label{rebiddingbycapa}}{\includegraphics[width=0.45\linewidth]{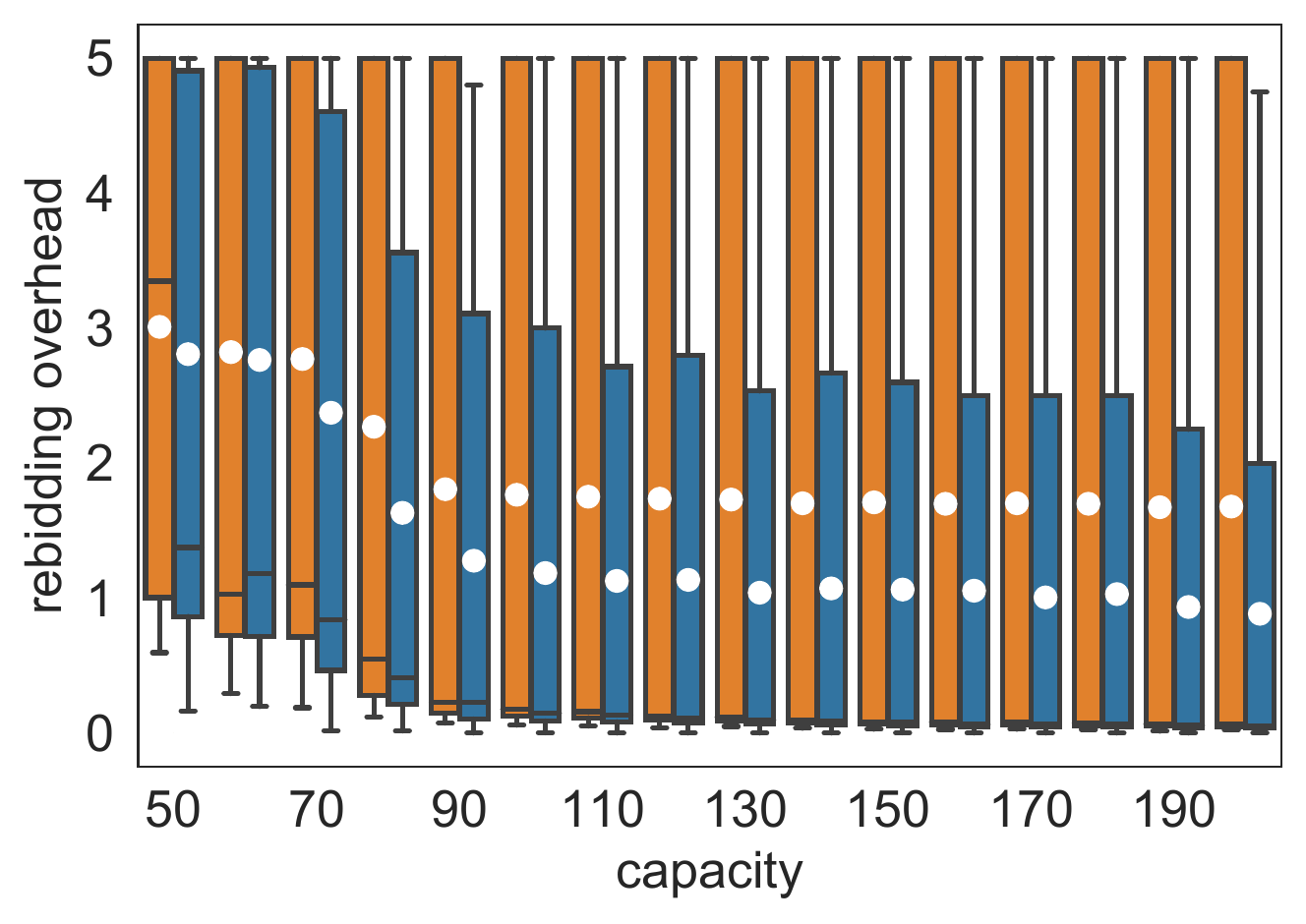}}\hspace{1em}
	\subcaptionbox{Remote site resource utilization comparison with MP=1. M+R\_1 utilizes resource by $18\%$ more than RIAL in high contention. In low contention, capacity is less critical, utilization is similar. M+R\_1 reduces the standard deviation in utilization by up to $21\%$\label{utilizationbycapa}}{\includegraphics[width=0.45\linewidth]{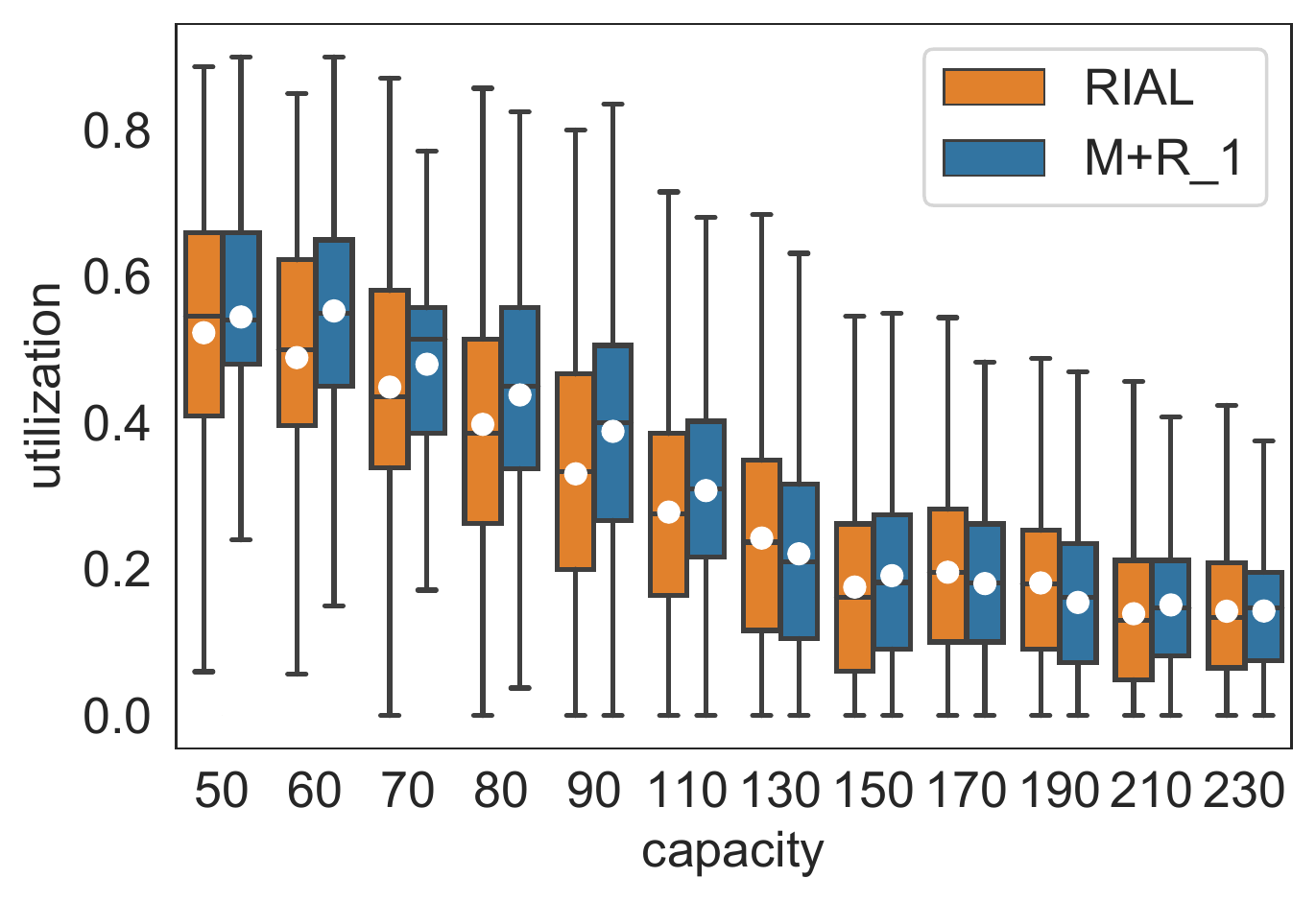}}
	\vspace*{-0.2cm}
	\caption{Performance comparison between: MALFOY+RIAL with immediate reward within 1 time step (M+R\_1), and only RIAL, with maximum permitted rebidding (MP) of 1 and 5 times. (a): offloading failure rate (OFR) vs system resource capacity, (b): required capacity to reach given OFR requirement, (c): rebidding overhead, (d): resource utilization in varying levels of capacity}
	\label{performancebycapa}
\end{figure*}

We test our approach in two steps. First, we comprehensively study the performance of active agents with MALFOY algorithm in a synthetic setup with a reward signal that becomes available to the agents at the end of every time step---this setup is the same as in \cite{tan2022multi}, which is a special case of long-term reward maximization with reward interval of $1$ time step (denoted M+R\_1). In this setup, we simplify the modeling of communication channel and vehicle mobility and focus on analyzing the effect of environmental parameters to the learning process---system resource capacity, maximum permitted number of rebiddings, and a large number of different service types.

Next, we use a realistic setup with a delayed extrinsic reward signal every $2000$ time steps (denoted M+R\_2000) to train and evaluate our model, showing the long-term effects of learning in the training environment. A generalizable model should be able to run in a different test environment without retraining and still achieve good performance. Therefore, to demonstrate our model's generalization properties, we initialize the agents with trained models from the training environment, and we run them again in the test environment without retraining. The two environments differ in number of vehicles, speed, arrival rate, traffic light phases, and system resource capacity; details are in Sec. \ref{subsec:eval_real} and Fig. \ref{generalization}.

The intrinsic reward signal includes immediate bidding outcome, payment, and system resource utilization; the extrinsic reward is the vehicle's cumulated gain from repeated auctions since the previous reward signal. In this setup, we keep the environmental parameters constant and model a 4-way traffic intersection, data transmission delay and vehicle mobility (i.e., speed, number of vehicles in range, traffic light phases, etc.).

\subsection{Synthetic setup}
\label{subsec:eval_hypo}

In this setup, we cover a wide range of hypothetical scenarios by varying parameters such as system capacity, service/task types and number of rebidding. \begin{inparaenum}[1)] 
\item Task types by resource needs in time-resource units: F1: 3 units, and F2: 30 units. We assume that tasks can be executed on multiple CPUs, such that the processing time is the reciprocal of the resource amount allocated, and the product of processing time and resource amount is the constant value of resource needs. A simplification is the assumption of independence between two types of resources. If the duration calculated from the allocation of two resource types are different, we take the longer duration as the processing time.
\item Service types by deadline and probability: F1, $300$ms: $18.75\%$; F1, $50$ms: $18.75\%$; F2, $300$ms: $6.25\%$; F2, $50$ms: $6.25\%$; F1-F2, $300$ms: $18.75\%$; F1-F2, $50$ms: $18.75\%$; F2-F1, $300$ms: $6.25\%$; F2-F1, $50$ms: $18.75\%$. We predefine the distribution from which the service types are drawn. More detailed analysis of these hyperparameters is left to future work.
\item Service arrival rate per vehicle: randomized according to the MMPP, with our predefined parameters $\lambda_\text{high} \in (0.48,0.6), \lambda_\text{low} \in (0,0.12)$ and transition probabilities $p_\text{high}=p_\text{low}=0.6$. More detailed analysis of these hyperparameters is left to future work.
\item Capacity: $50$-$230$ resource units. 
\item Maximum permitted rebidding: $1$ or $5$ times, respectively. 
\item Vehicle count: constant at $30$. 
\item Vehicle arrival rate: $0$, always in the system; speed: $0$. 
\item Data size: uniform random between $2.4$-$9.6$kbit. 
\item Uplink and downlink latency: $0$.
\item Extrinsic reward signal interval: $1$ time step.
\end{inparaenum}

	\begin{figure}[t]
		\centering
		\begin{minipage}{\linewidth}
		\centering
		\subcaptionbox{CDF of individual OFRs (capacity=70, MP=1): M+R\_1 does not sacrifice individual OFR to improve system performance: the CDF curves of each vehicle's OFR moves to the left in both budget categories. M+R\_1 is also fairer: vehicles with low budget reduced failure rate more than those with high budget.  \label{cdf}}{\includegraphics[width=0.9\linewidth]{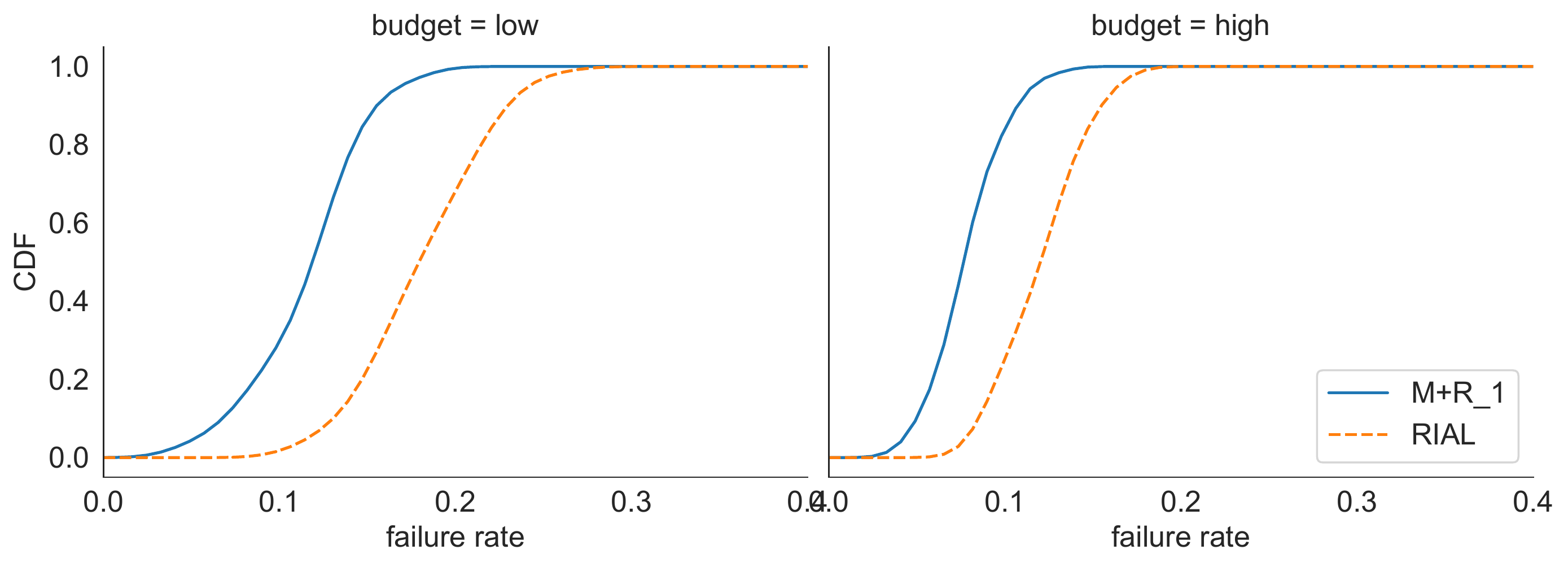}}\hfill
		\subcaptionbox{Tradeoff between backoff and bidding price, for tasks with long and short deadlines: an example with capacity=50, MP=5. Vehicles that bid low (high) use long (short) backoff. Vehicles learn to utilize backoff to overcome budget disadvantage. \label{backoffprice}}{\includegraphics[width=0.45\linewidth]{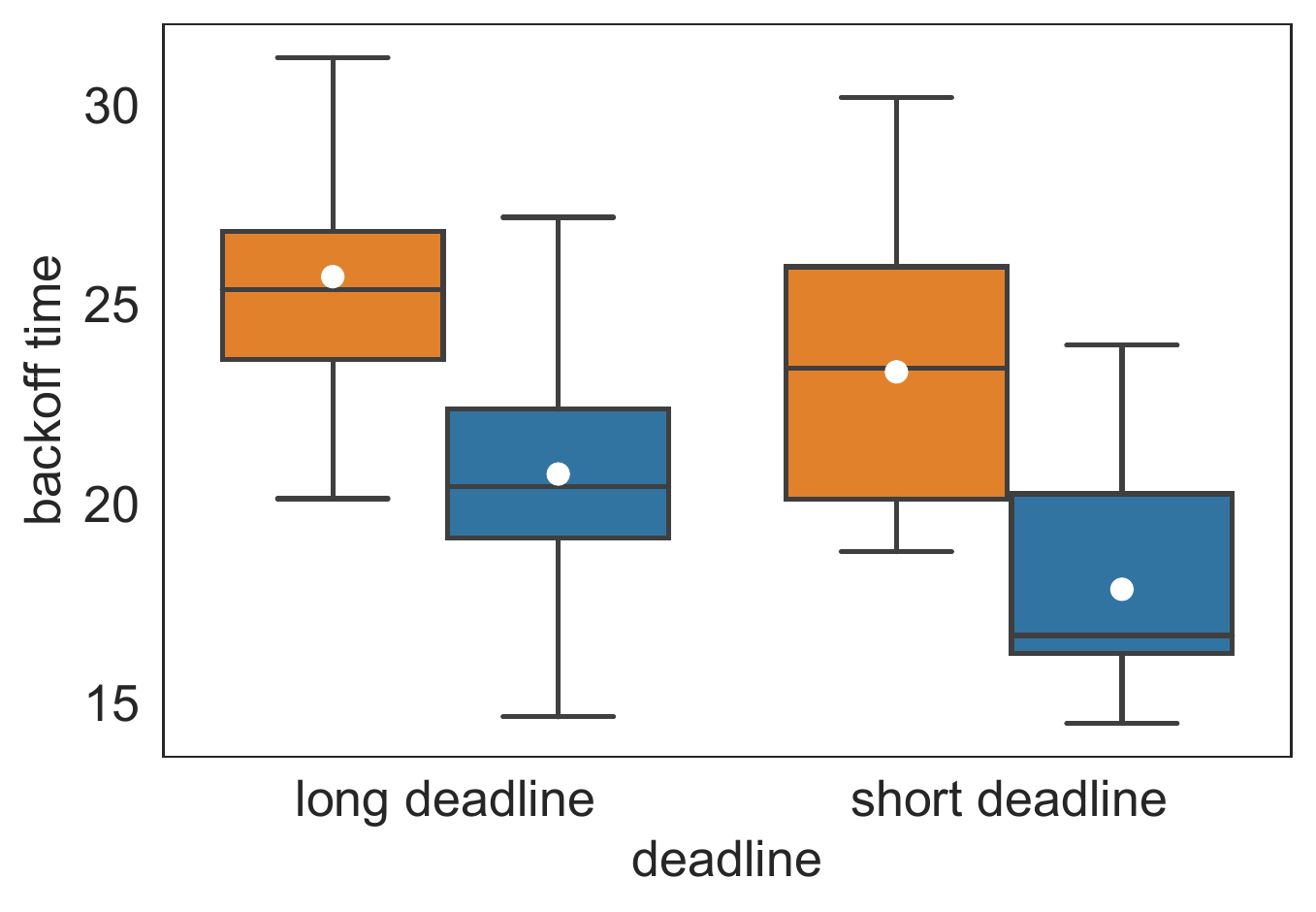}}\hspace{1em}
		\subcaptionbox{The same backoff-price tradeoff in different capacity levels, an example with MP=5 and tasks with long deadline: backoff time decreases as capacity increases, but the tradeoff effect remains.\label{backoffpricebycapa}}{\includegraphics[width=0.45\linewidth]{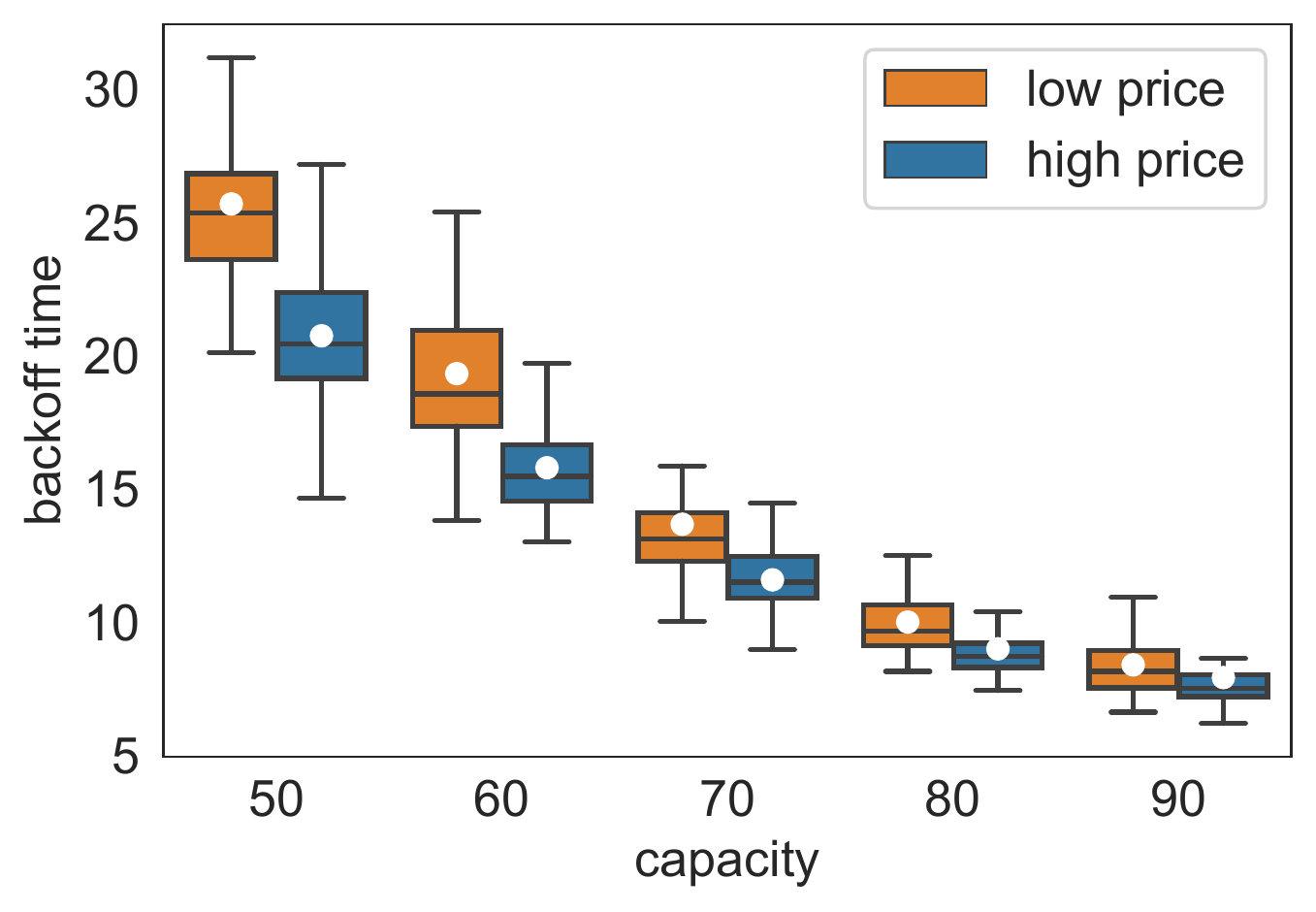}}
		\end{minipage}
		\vspace*{-0.2cm}
		\caption{Cumulative Distribution Function (CDF) of individual vehicles' offloading failure rates (OFR), backoff and price tradeoff}
		\vspace*{-0.2cm}
		\label{backoff}
	\end{figure}

As demonstrated in Figures \ref{successbycapa} and \ref{capause}, our active agents adapted to an environment with delayed information and learned to better utilize computing site resources. Fig.\ref{utilizationbycapa} shows how M+R\_1 increases computing site utilization in high contention and reduces load variation. When more rebidding is permitted, low OFR can be achieved by trial-and-error, and the advantage of MALFOY's backoff strategy is limited. That is why higher MP reduces M+R's advantage over RIAL. However, trial-and-error comes with a cost: Fig.\ref{rebiddingbycapa} compares the rebidding overhead used by both algorithms when MP=$5$. In high contention, both active and passive agents leverage on rebidding, and the difference in rebidding overhead is small. M+R's advantage becomes more significant as capacity increases.  

Fig.\ref{cdf} shows the cumulative probability of vehicles' individual OFRs. With MALFOY, as system overall OFR reduces, the individual OFRs reduce accordingly: the auction does not cause disadvantage to individual bidders. Moreover, vehicles with lower budget improve by a greater margin: they learn to utilize backoff mechanism to overcome their disadvantage in initial parameterization. Fig.\ref{backoffprice} shows how vehicles learn to trade off between bidding price and backoff time. They are separated into two groups: a vehicle is in the ``low price'' group if it bids on average lower than the average bidding price of all vehicles; otherwise, it is in the ``high price'' group (here we analyze actual bidding prices instead of the predefined budgets). When service requests have a longer deadline, vehicles in both price groups learn to utilize longer backoff. ``low price'' vehicles always use longer backoff. Fig.\ref{backoffpricebycapa} shows tradeoff is present in all capacity levels. 

To summarize: Fig.\ref{performancebycapa} demonstrate MALFOY's excellent overall system performance; Fig.\ref{cdf} shows that system objective is aligned with individual objectives through incentivization (\textbf{C1}), and especially in Fig.\ref{backoff}, differently initialized agents learn to select the most advantageous strategy based on limited feedback signal (\textbf{C2}). The capability to learn and behave accordingly makes our agents highly flexible in a dynamic environment. 

\subsection{Realistic setup}
\label{subsec:eval_real}

	\begin{figure*}[t]
		\centering
		\begin{minipage}{\linewidth}
		\centering
		\subcaptionbox{Training environment: low contention with abundant resource, traffic phase=$10$-$40$s, low vehicle speed($10$km/h), low arrival rate=($1/2.2\text{s}$), low variation in vehicle count($22$-$30$): OFR in training(left) and evaluation(right). M+R\_2000 with long-term objective learns faster in training, and outperforms both M+R\_1 with only short-term objective, and RIAL. \label{general-train}}
			 			{ \includegraphics[width=0.45\linewidth]{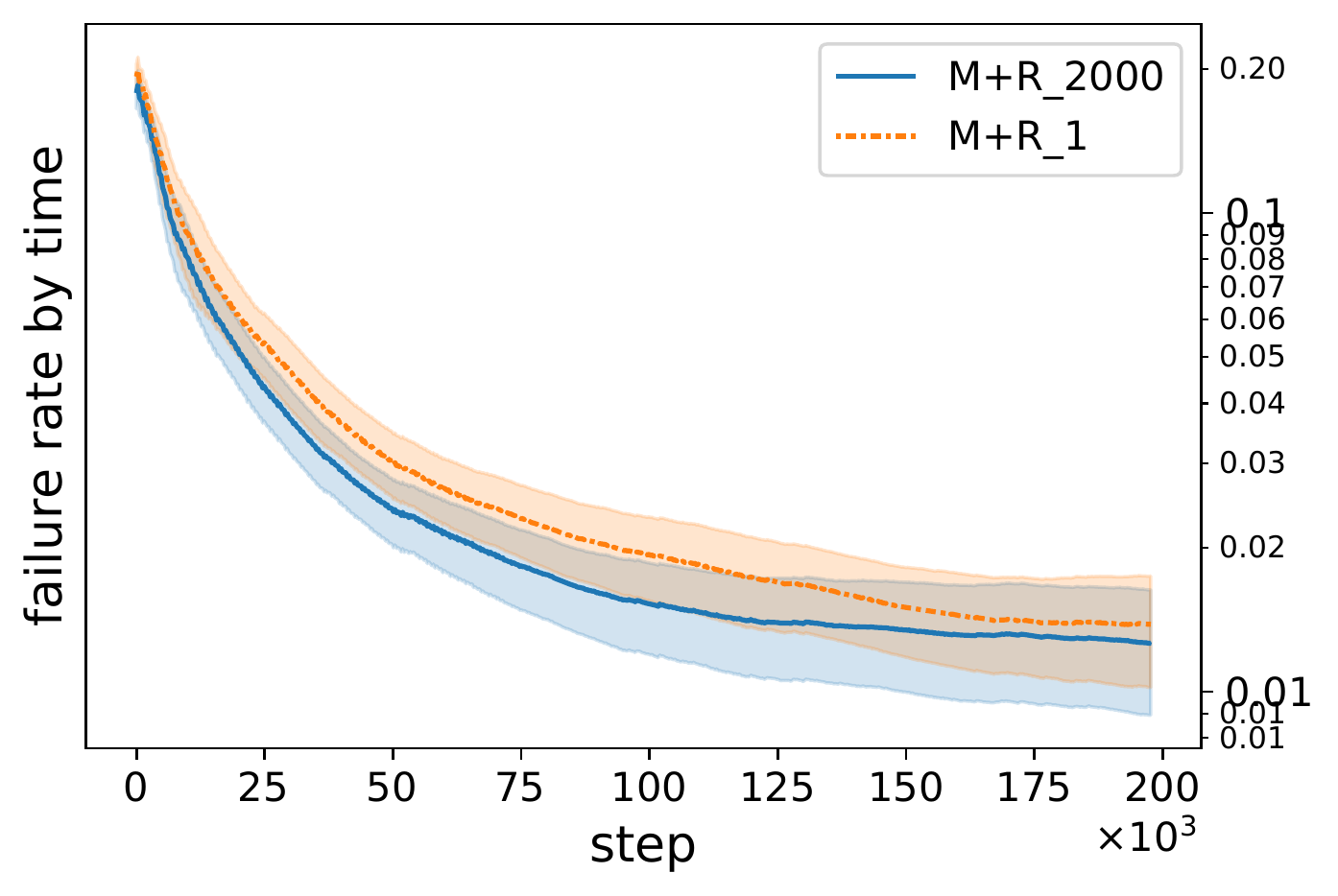}\hfill
						   \includegraphics[width=0.45\linewidth]{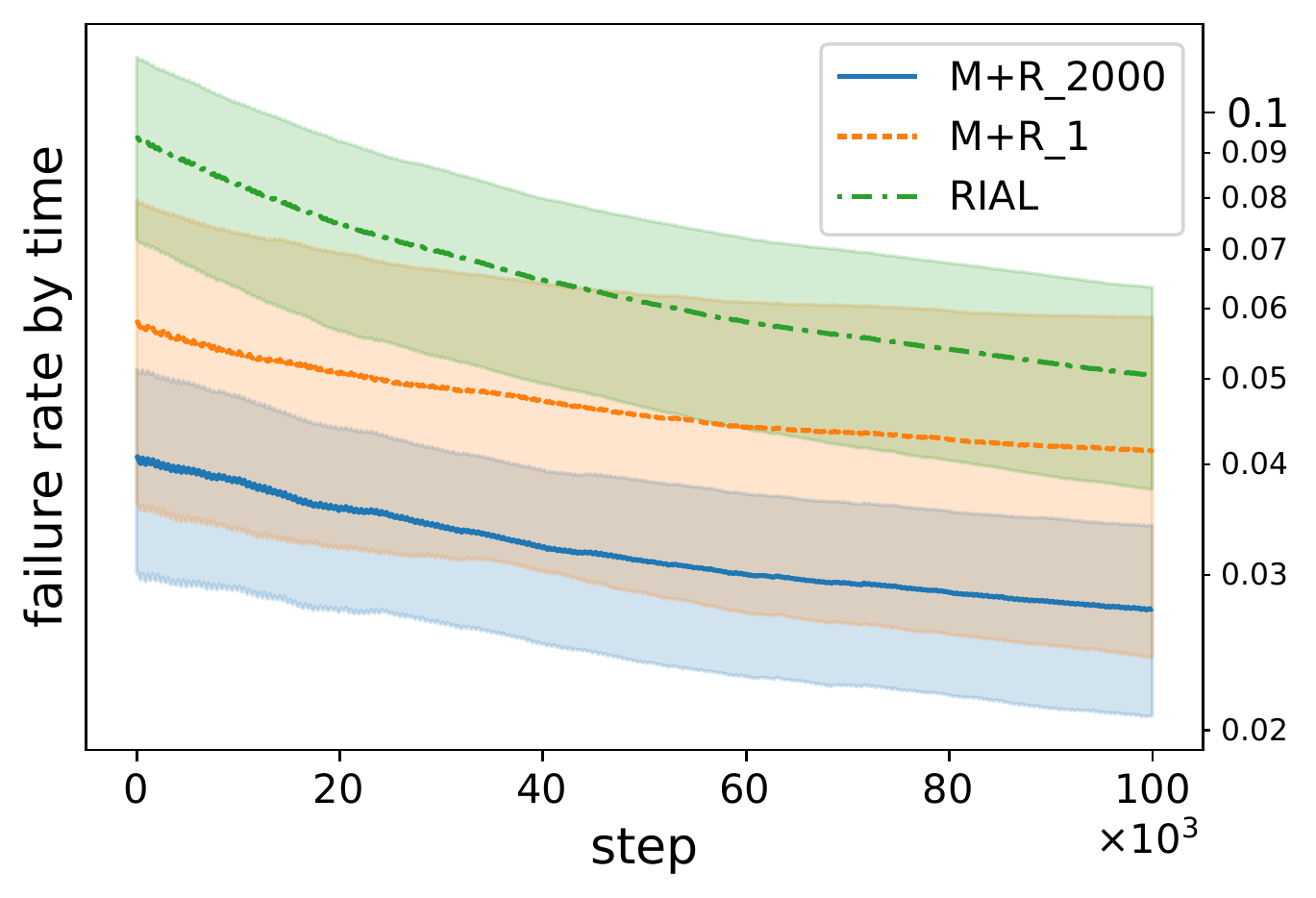}
						}\hfill
		\subcaptionbox{Test environment: high contention with limited resource, traffic phase=$20$s, high vehicle speed($30$km/h), high arrival rate($1/1\text{s}$), high variation in vehicle count($14$-$30$): vehicle count over time(left) and OFR(right). Vehicle count shows the volatility of the test environment. OFR performance of M+R\_2000 with long-term objective is even more distinguishable from M+R\_1 and RIAL. \label{general-test2}}
						{   \includegraphics[width=0.45\linewidth]{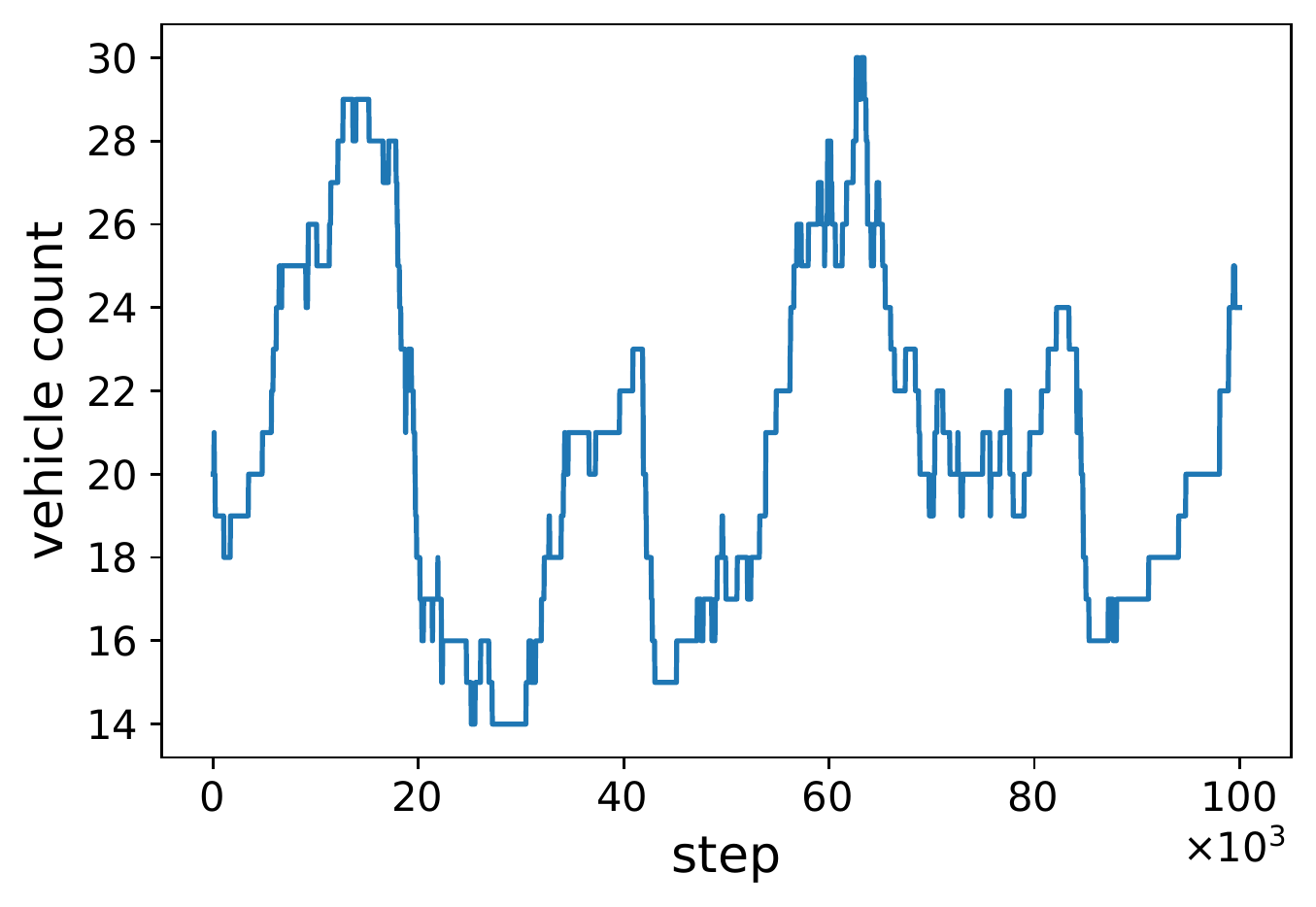}\hfill
						    \includegraphics[width=0.45\linewidth]{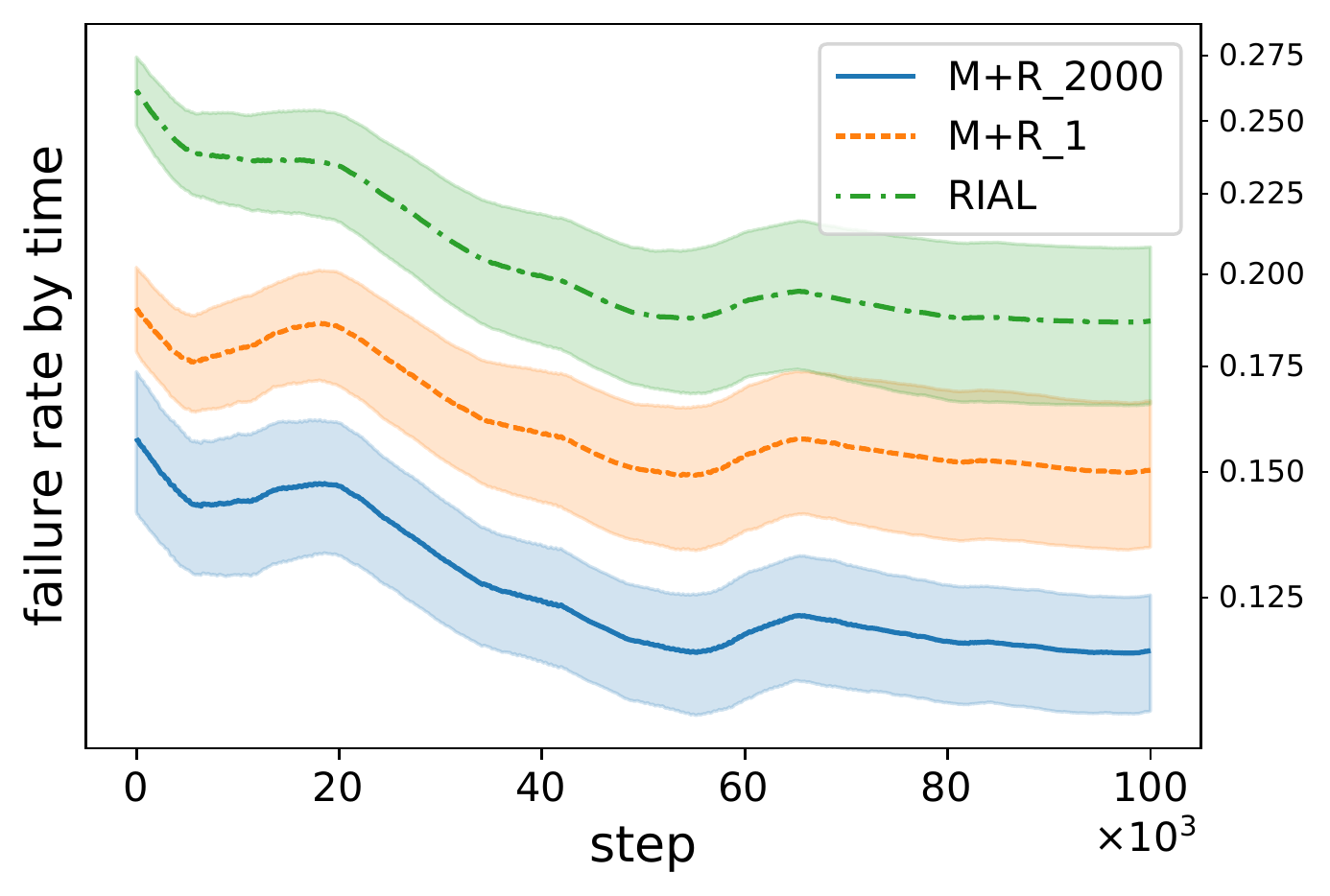}
						}\hfill
		\end{minipage}
		\vspace*{-0.2cm}
		\caption{Comparison of offloading failure rate (OFR) in training and test environments, between: MALFOY+RIAL with external reward delay of 2000 time steps (M+R\_2000), MALFOY+RIAL with immediate reward (M+R\_1), and RIAL only.}
		\label{generalization}
	\end{figure*}

In this setup, we adopt the data patterns of segmentation and motion planning applications extracted from various self-driving data projects \cite{cordts2016cityscapes} and referenced from relevant studies \cite{chen2017importance,broggi2014proud}. We also use Simulation of Urban Mobility (SUMO) \cite{behrisch2011sumo} to create a more realistic mobility model of a single junction with a centered traffic light. Information of the junction is downloaded from open street map. Assuming 802.11ac protocol, we place the ACA unit in the middle of the graph and limit the edges to within 65m of the ACA unit. The net is with two lanes per street per direction, SUMO uniform-randomly creates a vehicle at any one of the four edges. Also, in the realistic setup, we consider a sparse and delayed reward signal with an interval of $2000$ time steps.

Parameters of the setup are as follows \cite{cordts2016cityscapes,chen2017importance,broggi2014proud}: \begin{inparaenum}[1)]
\item Task types: F1: $80$ units, and F2: $80$ units. 
\item Service types and deadline: F1: $100$ms and F2: $500$ms. 
\item Service arrival rate per vehicle: fixed at F1: every $100$ms, and F2: every $500$ms. 
\item Capacity: $20$ in high contention, $30$ in low contention. 
\item Maximum permitted rebidding: $1$. 
\item Vehicle count: $14$-$30$ from simulated trace data. 
\item Vehicle arrival rate: constantly at $1$ every $1$ or $2.2$ seconds; speed: $10$ or $30$ km/h when driving. 
\item Data size: uplink: F1: $0.4$Mbit, F2: $4$Mbit. Downlink: F1: $0$ (negligible), F2: $0.4$Mbit. 
\item Latency: we take 802.11ac protocol that covers a radius of 65 meters, and assume maximum channel width of ca. $1.69$ Gbps. We model the throughput as a function of distance to the ACA unit: throughput=$-26 \times \text{distance} + 1690$ Mbps \cite{shah2015throughput}. If there are $N$ vehicles transmitting data to the ACA unit, we assume that each gets $1/N$ of the maximum throughput at that distance.
\item Extrinsic reward signal interval: $1$ or $2000$ time steps.
\end{inparaenum}

As mentioned in Sec.\ref{subsubsec:servicerequest}, the uplink and downlink time, service request arrival rate and service deadlines are based on the requirements of semantic segmentation and motion planning applications. If the vehicle expects its position before service deadline to be out-of-range of the MEC, the service request is dropped without any performance measurement.

Higher vehicle arrival rate and slower driving speed typically lead to high contention. By changing the arrival rate and speed in the simulation, we create high and low-contention scenarios alternatively. 

For training, we set the traffic light phases to $10$-$40$s of green for each direction, alternatively. We train and test our active agents with MALFOY in low contention, with reward signal interval at $1$ and $2000$ time steps, denoted M+R\_1 and M+R\_2000. Fig.\ref{general-train}-left shows that M+R\_1 converges to OFR of $1.4$\%, and M+R\_2000 converges much faster to an even lower failure rate. Then we evaluate the trained models in the same environment with newly simulated trace data from SUMO (Fig.\ref{general-train}-right), M+R\_1 still reaches OFR of $4$\%, a reduction of $18\%$ compared to RIAL; M+R\_2000 further reduces failure rate by $34\%$, compared to M+R\_1.

	\begin{figure}[t]
		\centering
		\begin{minipage}{0.95\textwidth}
         \centering
		\includegraphics[width=0.6\textwidth]{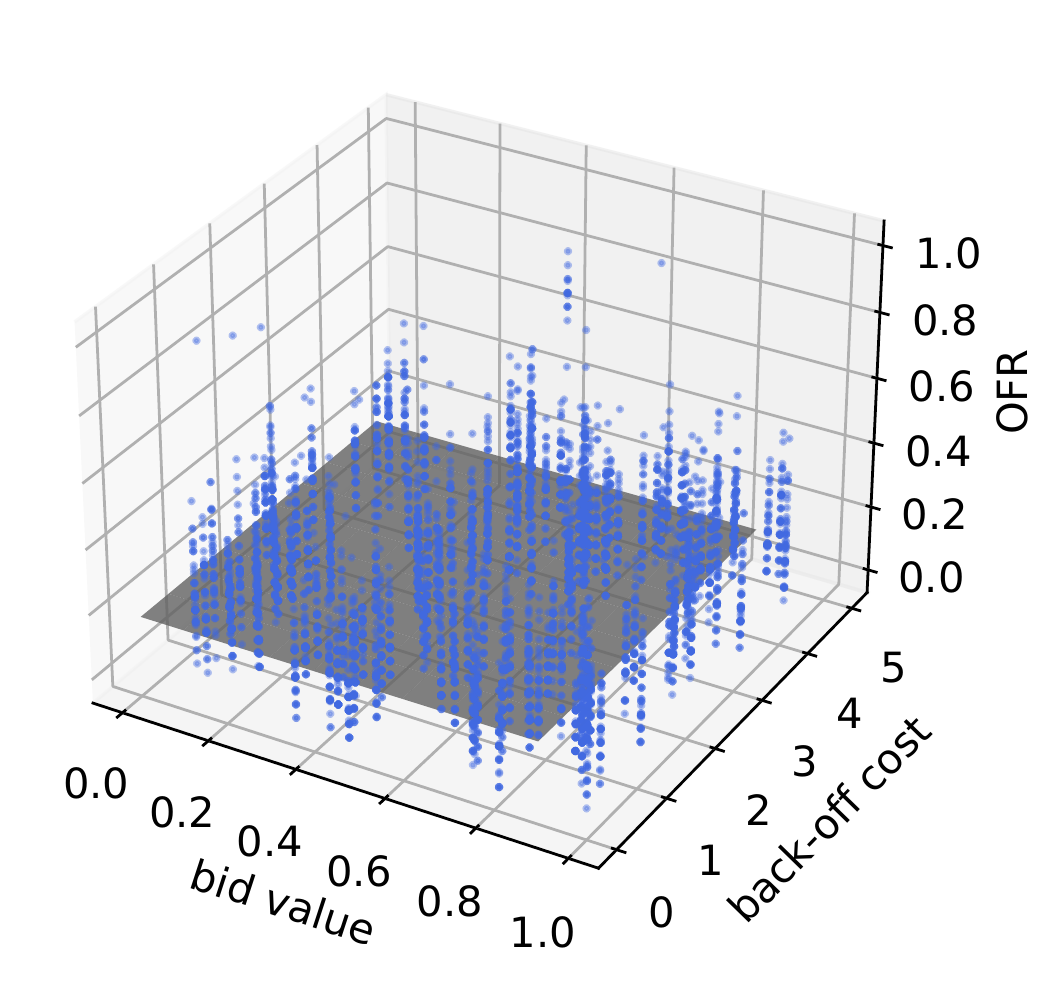}\hfill
		\end{minipage}
		\vspace{-0.2cm}
		\caption{Vehicles' individual OFR is not sensitive to private bid values $v$ and back-off cost $q$ (normalized).}
		\label{sensitivity}
	\end{figure}

Then, we test (i.e.\ without retraining) the trained MALFOY models in a significantly different environment, changing traffic light phases, vehicle arrival rate and speed to make the environment more volatile and dynamic, and reducing capacity to create a high-contention situation. The resulting vehicle count over time (Fig.\ref{general-test2}-left) shows a much heavier and more frequent fluctuation compared to the original training environment. Note that vehicle count and OFR do not vary synchronously---OFR is determined by vehicle count and numerous other complicating factors such as transmission, queueing and processing time, past utilization, etc. Despite the significant changes to the environment, and without requiring any further training, M+R\_1 reduces failure rate by $20\%$ compared to RIAL, and M+R\_2000 further reduces failure rate by $23\%$ (Fig.\ref{general-test2}-right). 

Fig.\ref{general-train} shows good convergence speed despite computation and communication complexity of the problem (\textbf{C3}). Fig.\ref{general-test2} shows that MALFOY has very good generalization properties---in fact, in the more volatile and dynamic environment, the superiority of active agents becomes more obvious. With the capability to predict long-term impacts of each action, MALFOY shows even better performance and generalization properties (\textbf{C4}). With little need for retraining in a new environment, the computation delay is only the time for model inference. Test of inference time on a vehicle OBU is partially dependent on the hardware, therefore it is not the scope of this study.

Additionally, we randomize each vehicle's private bid values $v$ and the back-off cost $q$, to analyze how sensitive the individual offloading failure rate (OFR) is to changes in $v$ and $q$. Results show that the changes in $v$ and $q$ have almost no impact on the individual OFR; the Pearson coefficient values are $0.008$ (p-value=$0.5$) and $0.007$ (p-value=$0.5$), respectively. Fig. \ref{sensitivity} visualizes this result. This and the results in Fig. \ref{backoff} demonstrate the robustness of our auction mechanism: vehicles learn to compensate for differences in initial parameterization through trade-off in bidding price and backoff time, without impact on individual OFR.

To summarize: results in the synthetic setup show that, compared to only having a centralized load-balancing solution at the MEC, MALFOY succeeds in incentivizing each autonomous vehicle to add to the load-balancing effect in a distributed manner, which significantly increases resource utilization in high contention, and reduces capacity needed to reach the same service level. It achieves this by letting each vehicle independently decide how to trade off between backoff time and bidding price. Results in the realistic setup shows MALFOY's excellent generalization property in different realistic environments, making it a potential add-on to any existing centralized solutions at the MEC. A sensitivity analysis shows the robustness of our solution.

\section{Conclusion}
\label{sec:conclusion}

Our agents learn how to best utilize backoff option based on its initialization parameters. As a result, the agents achieve significant performance gains in very different environments. MALFOY can utilize long-term, sparse reward signals and has enhanced predictive power, as well as better alignment between short-term and long-term goals. When behaving long-term, it shows further performance improvements. Our interaction mechanism aligns private and system goals without sacrificing either user autonomy or system-wide resource efficiency, despite the distributed design with limited information-sharing.

The algorithm is therefore applicable to a wide range of distributed resource allocation problems in a dynamic and adversarial environment with no or very limited \textit{a priori} information, large number of autonomous users with private goals, and large number of custom service requests. We find such applications in e.g., telecommunications, energy, Internet of Things, vehicular networks, cloud computing, etc.

In this paper, we fix the hyperparameters of the algorithm for the simulation, such as penalty costs related to backoff decisions and lost bids, each agent's preferences of long and short-term objectives, etc. A meta-learning algorithm that learns the best hyperparameters is left to future work. Besides, we assume there is no ``malicious'' agent with the goal to reduce social welfare or attack the system.

\section{Appendix}
\label{sec:appendix}

\subsection{Summary: theoretical results}
\subsubsection{Low contention}
\label{lowContention}

We show that in low contention, the interaction mechanism is a potential game with NE. We use the concept of potential functions to do so \cite{monderer1996potential}:

\begin{defi} $G(I,A,u)$ is an exact potential game if and only if there exists a potential function $\phi(A): A \to \mathbb{R}$ s.t. $\forall i \in I$, $u_i(b_i,b_{-i})-u_i(b'_i,b_{-i})=\phi_i(b_i,b_{-i})-\phi_i(b'_i,b_{-i}), b \in A$.
\end{defi} 

\begin{rem}\label{potentialNE} Players in a finite potential game that jointly maximize a potential function end up in NE. \end{rem}

\begin{proof} See \cite{monderer1996potential}. \end{proof}

\begin{thm} Bidders with utility as Eq.\ref{eq:reward1} participate in a game as described in Sec.\ref{sec:problem} in low contention, the game is a potential game, and the outcome is an NE.\end{thm}

\begin{proof}
In low contention, $p_{i,k}=0$, as all bids are accepted. $u_i$ is reduced to: $u_i(\alpha_i,\alpha_{-i})=\sum\limits_{k}q_{i,k}-\sum\limits_{k}\alpha_{i,k} q_{i,k} + W \Big(1-\sum_j \alpha_j \cdot \frac{\omega_j}{C}\Big)$, where $-i$ denotes bidders other than $i$. $\omega_j \in \mathbb{R}^{|K|}$ is each bid's resource requirement, $C$ is system capacity. Thus, the auction is reduced to a potential game with discrete action space $\alpha_i \in \mathbb{R}^{|K|}$, and potential function $\phi(\alpha_i,\alpha_{-i})=\sum\limits_{j, k}q_{j,k}-\sum\limits_{j,k} \alpha_{j,k}q_{j,k} + W\Big(1-\sum_j \alpha_j \cdot \frac{\omega_j}{C} \Big), \forall i,j \in I, \forall k \in K$. 

We prove in Appendix \ref{appendix:potentialGame} that $u_i(\alpha_i,\alpha_{-i})-u_i(\alpha'_i,\alpha_{-i})=\phi(\alpha_i,\alpha_{-i})-\phi(\alpha'_i,\alpha_{-i})$, and hence it is a potential game, and bidders maximizing their utilities $u_i$ also maximize the potential function $\phi$. Since $\alpha_i \in \mathbb R^{|K|}$, it is a finite potential game. According to Remark \ref{potentialNE}, the outcome is an NE.
\end{proof}

In low contention, our computation offloading problem becomes a potential game. This enables us to use online learning algorithms such as in \cite{perkins2014game} that converge regardless of other bidders' behaviors. The NE is a local maximization of the potential function: each bidder finds a balance between its backoff cost and the incentive to reduce overall utilization. Empirical results in Sec.\ref{sec:eval} confirm that over time this results in a more balanced load.

\subsubsection{High contention}
\label{highContention}

In high contention, $\alpha$ is used in a repeated auction to avoid congestion and ensure better reward over time. To simplify the proofs, we consider only the time steps where $\alpha=1$ (bidder joins auction). We also take a small enough $W$, such that the last term in Eq.\ref{eq:reward1} can be omitted in high contention, to further simplify the utility function in the proof.

\begin{thm}\label{thm:spa} In a second-price auction, where bidders with utility as Eq.\ref{eq:reward1} compete for service slots as commodities in high contention, \begin{inparaenum}[1)] \item bidders' best-response is of linear form, \item the outcome is an NE and \item welfare is maximized.\end{inparaenum}
\end{thm}

\begin{proof} See Appendix \ref{appendix:SPAwithpenalty}.\end{proof} 

When bidders bid for service slots, the required resources are allocated. Theorem \ref{thm:spa} guarantees the maximization of welfare (total utility of bidders), but it does not guarantee the optimality of the resource allocation, unless the following conditions are met: if bidders' valuation of the commodity is linear to its resource requirement, and all bidders have some access to resources (fairness). 

\begin{cor}\label{pareto} In a second-price auction, where $M$ bidders with utility as Eq.\ref{eq:reward1} compete in high contention, the outcome is an optimal resource allocation, if the bidders' valuation of commodities is linear to resource requirement and all bidders have a positive probability of winning.
\end{cor}

\begin{proof} See Appendix \ref{appendix:paretoOptimal}.\end{proof} 

Our setup meets both conditions.

\subsection{Proof of potential game}
\label{appendix:potentialGame}

\begin{proof} 

We define player $i$'s utility as $u_i(\alpha_i,\alpha_{-i})=\sum\limits_{k \in K}q_{i,k}-\sum\limits_{k \in K}\alpha_{i,k} q_{i,k} + W \Big(1-\frac{\sum_j \alpha_j \cdot \omega_j}{C}\Big)$, where $\omega_j \in \mathbb{R}^K$ is the resource requirement of each commodity, $C$ is the system capacity. 

We define potential function: $\phi(\alpha_i,\alpha_{-i})=\sum\limits_{j \in I, k \in K}q_{j,k}-\sum\limits_{j \in I, k \in K} \alpha_{j,k}q_{j,k}+W\Big(1-\frac{\sum_j \alpha_j \cdot \omega_j}{C} \Big)$. 

To simplify, we substitute with $Q_i=\sum\limits_{k\in K}q_{i,k}$, $A_i=\sum\limits_{k \in K}\alpha_{i,k} q_{i,k}$, $A_{-i} = \sum\limits_{j \in I, j \neq i, k \in K}\alpha_{j,k} q_{j,k}$, $B_i=\sum\limits_k \alpha_{i,k} \omega_{i,k}$, $B_{-i}=\sum\limits_{j \in I, j \neq i, k \in K}\alpha_{j,k} \omega_{j,k}$, and rewrite: $u_i(\alpha_i,\alpha_{-i})=Q_i-A_i+W-\frac{W}{C}(B_i+B_{-i})$, $u_i(\alpha'_i,\alpha_{-i})=Q_i-A'_i+W-\frac{W}{C}(B'_i+B_{-i})$, $\phi(\alpha_i,\alpha_{-i}) = \sum\limits_j Q_j-(A_i+A_{-i})+W-\frac{W(B_i+B_{-i})}{C}$, $\phi(\alpha'_i,\alpha_{-i}) = \sum\limits_j Q_j-(A'_i+A_{-i})+W-\frac{W(B'_i+B_{-i}) }{C} \implies u_i(\alpha_i,\alpha_{-i})-u_i(\alpha'_i,\alpha_{-i}) =-(A_i-A'_i)-\frac{W}{C}(B_i-B'_i) =\phi(\alpha_i,\alpha_{-i})-\phi(\alpha'_i,\alpha_{-i})$
\end{proof}

Since $\alpha_i \in \mathbb R^{|K|}$, the game under low contention is a finite potential game.

\subsection{Second-price auction}
\label{appendix:SPAwithpenalty}

Under high contention, as defined in Sec.\ref{payment}, $u_i$ is reduced to: \begin{flalign}\label{eq:ui}
u_i= \sum\limits_{k \in K} \Big(z_{i,k} \cdot (v_{i,k}-p_{i,k})-(1-z_{i,k}) \cdot c_{i,k} \Big)
\end{flalign}

We prove the theorem for $|M|=2$ and $|K|=1$. It is an extension from \cite{sun2006wireless}. Unlike \cite{sun2006wireless}, we include in utility definition the second-price payment and cost for losing a bid. Based on \cite{sun2006wireless}, it can also be easily extended to multiple bidders. 

\subsubsection{Basic model}

$2$ bidders receive continuously distributed valuations $v_i \in [l_i,m_i], i \in\{1,2\}$ for $1$ commodity, and choose  their strategies $f_1(v_1),f_2(v_2)$ from the strategy sets $F_1$ and $F_2$. The resulting NE strategy pair is $(f_1^*, f_2^*)$. Any strategy function $f(v)$ is increasing in $v$, with $f_1(l_1)=a$, and $f_1(m_1)=b$. We also assume the users have budgets $(B_1,B_2)$, and that they cannot bid more than the budget. We define cost for losing the bid $c_i$.
Furthermore, we define the inverse function of $f_1(v_1)$ to be: $h_1(y_1)= l \text{, if } y_1\leq a_1 \text{, } h_1(y_1)=f_1^{-1}(y_1) \text{, if } a_1<y_1<b_1 \text{, and } h_1(y_1)=m \text{, if }y_1 \geq b_1$.

For a given $f_1$, if bidder 2 chooses a bidding function $f_2$, according to Eq.~\ref{eq:ui}, the expected utility for bidder 2 is $u_2(f_1,f_2)=\mathbb{E}_{v_1,v_2}[(v_2+c_2) \cdot 1_{f_2(v_2)\geq f_1(v_1)}] - \mathbb{E}_{v_1,v_2}[f_1(v_1)\cdot 1_{f_2(v_2)\geq f_1(v_1)}]-c_2$, where $1_{f_2(v_2)\geq f_1(v_1)}=1 \text {, if } f_2(v_2)\geq f_1(v_1) \text{, otherwise } 0$. To simplify, we define $E_1 =\mathbb{E}_{v_1,v_2}[(v_2+c_2) \cdot 1_{f_2(v_2)\geq f_1(v_1)}]$ and $E_2=\mathbb{E}_{v_1,v_2}[f_1(v_1)\cdot 1_{f_2(v_2)\geq f_1(v_1)}]$. Hence, $u_2(f_1,f_2)=E_1-E_2-c_2$. $E_2$ is the expected second price payment when bidder 2 wins, and the payment should be no greater than $\min(b_2,B_2)$. Since to avoid overbidding, we assume $b_2 \leq B_2$, the set of feasible bidding functions for bidder 2 given $f_1$ is $S_2(f_1) = \{ f_2 \in F_2|u_2(f_1,f_2) \geq 0,E_2 \leq b_2 \}$.

For the condition $u_2(f_1,f_2)\geq 0$ to hold, we can prove that at any point where  $1_{f_2(v_2)\geq f_1(v_1)}=1$, we have $v_2 \geq f_1(v_1)$, which is a sufficient condition of $u_2(f_1,f_2)\geq 0$. This is because $f_2$ is bidder 2's bidding signal, to avoid overbidding, $f_2(v_2) \leq \min(b_2, v_2)$, therefore $f_1(v_1) \leq v_2$. We thus simplify the above equation to: $S_2(f_1) = \{ f_2 \in F_2| E_2 \leq b_2 \}$.

We formulate the problem into a utility maximization problem: $\max\limits_{f_2 \in S_2(f_1)} u_2(f_1,f_2)$. We say $f_2$ is a best response of bidder 2, if $u_2(f_1,f_2)\geq u_2(f_1,f_2')$, $\forall f_2' \in S_2(f_1)$. A NE strategy pair $(f_1^*, f_2^*)$ has the selected strategies as each other's best responses. 

\subsubsection{Form of the best response}

\begin{thm}\label{thm:bestResp} Given bidder 1's bidding strategy $f_1 \in F_1,f_1(l_1)=a_1,f_1(m_1)=b_1$, bidder 2's best response has the form $\begin{cases}
f_2(v_2) \leq a_1 & \text{for } v_2 \in [l_2,\theta_1] \\
f_2(v_2) = j_2 \cdot v_2 + d_2& \text{for } v_2 \in [\theta_1,\theta_2] \\
f_2(v_2) \geq b_1 & \text{for } v_2 \in [\theta_2,m_2]
\end{cases}$, where $\theta_1, \theta_2 \in [l_2,m_2]$ and $j_2 \theta_1 + d_2 =a_1, j_2\theta_2 + d_2 =b_1$.
\end{thm}

\begin{proof} Given $f_1$ and bidder 2's bid $y_2$, probability that bidder 2 wins the bid is:

$P_2^{win}(y_2)=P(f_1(v_1)\leq y_2)=P(v_1 \leq h_1(y_2)) =\int_{l_1}^{h_1(y_2)} \mathbf p_{1}(v_1)\mathrm{d}v_1$, where $\mathbf p$ is the probability density function, and $P$ is the cumulative function.

Bidder 2's optimization problem is: find a bidding function $y_2=f_2(v_2)$ to maximize $E_1-E_2$

\noindent$=\mathbb{E}_{v_1,v_2}[(v_2+c_2) \cdot 1_{f_2(v_2)\geq f_1(v_1)}] - \mathbb{E}_{v_1,v_2}[f_1(v_1)\cdot 1_{f_2(v_2)\geq f_1(v_1)}] =\int_{l_2}^{m_2}\int_{l_1}^{h(f_2(v_2))}\Big(v_2+c_2-f_1(v_1)\Big) \mathbf p_2(v_2) \mathbf p_1(v_1)\mathrm{d}v_1\mathrm{d}v_2$, s.t. $E_2 \leq b_2$.

To solve the optimization problem, we write the Lagrangian function with multiplier $\lambda$:

$\mathcal{L}(v_2,\lambda)=E_1-E_2 - \lambda (E_2 - b_2) = \int_{l_2}^{m_2} \Big[\int_{l_1}^{h_1(f_2(v_2))} V \mathbf p_1(v_1)\mathrm{d}v_1\Big] \mathbf p_2(v_2)\mathrm{d}v_2 - \lambda b_2$, where $V = v_2+c_2-(1+\lambda) f_1(v_1)$. 

Next, for each $v_2$, we find the $f_2$ that maximizes $\int_{l_1}^{h_1(y_2)}\Big(v_2+c_2-(1+\lambda)f_1(v_1)\Big)\mathbf p_1(v_1)\mathrm{d}v_1$, $y_2=f_2(v_2)$. $\max_{f_2}(\mathcal{L})$ is the equivalent of $\max_{f_2}(E_1)$.

For any given $v_2$, the above formula is the area below the function $\mathcal{z}=v_2+c_2-(1+\lambda)f_1(v_1)$, when $v_1$ moves in the range from $l_1$ to $h_1(y_2)$. As $f_1$ is monotonously increasing, $\mathcal{z}$ is monotonously decreasing. Therefore, to maximize the area below $\mathcal{z}$, $h_1(y_2)$ should simply be chosen as the intersection of $\mathcal{z}$ and the x-axis, or $v_2+c_2-(1+\lambda)f_1(h_1(y_2))=0$:

$y_2=f_1(f_1^{-1}(y_2))=f_2(v_2)=\frac{v_2+c_2}{1+\lambda}$, $\forall y_2\in [a_1,b_1]$, or $v_2 \in [(1+\lambda)a_1-c_2, (1+\lambda)b_1-c_2]$.

Since $f_2(v_2)$ is monotonously increasing, $f_2(v_2)\leq a_1, \text{ for } v_2 \in [l_2,(1+\lambda)a_1-c_2]$, and similarly, $f_2(v_2)\geq b_1, \text{ for } v_2 \in [(1+\lambda)b_1-c_2,m_2]$.

Theorem \ref{thm:bestResp} implies that the best response of bidder $1$ and $2$ are both of the linear form.
\end{proof}

\subsubsection{Existence of Nash equilibrium}
\label{appendix:ne}

\begin{thm}\label{thm:ne} When best response form is $f_1(v_1)=j_1v_1+d_1$ and $f_2(v_2)=j_2v_2+d_2$, we can always find a pair $(j_1,j_2)$ such that both bidders' budget range $[a_i,b_i]$ would be satisfied in NE. \end{thm}

\begin{proof}

A NE exists if there is a pair $(j_1,j_2)$ that satisfy the two constraints: $\mathbb{E}_{v_1,v_2}[f_1(v_1)\cdot 1_{f_2(v_2)\geq f_1(v_1)}] \leq b_2, \mathbb{E}_{v_1,v_2}[f_2(v_2)\cdot 1_{f_1(v_1)\geq f_2(v_2)}] \leq b_1$.

The following proves that such a pair exists. If we choose $c_1=c_2=c$, and given the linear best response forms, and given the bidders' bidding functions, we define $E_3=\mathbb{E}_{v_1,v_2}[(v_1-c)\cdot 1_{j_1v_1\geq j_2v_2}]$ and $E_4=\mathbb{E}_{v_1,v_2}[(j_2v_2-c)\cdot 1_{j_1v_1\geq j_2v_2}]$.

Define bidder 1's feasible strategy set: $S_1(j_2) = \{ j_1 \in [0,\infty) |  E_4 \leq b_1\}$. Due to its linear form, and according to Eq.~\ref{eq:ui}, bidder 1's best response is: $\mathbf b_1(j_2)=\argmax\limits_{f_1 \in S_1(j_2)} (E_3 -E_4) =\argmax\limits_{y \in S_1(j_2)} \mathbb{E}_{v_1,v_2}[v_1-j_2v_2\cdot 1_{y v_1\geq j_2v_2}]$. Utility $u_1(y) = \mathbb{E}_{v_1,v_2}[v_1-j_2v_2\cdot 1_{y v_1\geq j_2v_2}]$ is a non-decreasing function of $y$ defined on the set $S_1(j_2)$. To prove the existence of NE, we use Kakutani fixed point theorem.

\begin{thm}[Kakutani fixed point theorem \cite{ok2007real}]\label{kakutani} Let A be a non-empty, compact and convex subset of some Euclidean space $R^n$. Let $\varphi: A \to 2^B$ be an upper hemicontinuous set-valued function on A with the property that $\varphi(x)$ is non-empty, closed, and convex $\forall x \in A$. Then $\varphi$ has a fixed point. \end{thm}

We prove Lemmas \ref{lem1.1}-\ref{lem2.2} below, to show that our case meets the conditions of Theorem \ref{kakutani}. Hence, $\varphi: S_1 \to \mathbf b_1 \in 2^{S_1}$ has a fixed point, and there exists NE (Theorem \ref{thm:ne}).
\end{proof}

\begin{lem}\label{lem1.1} Bidder 1 strategy set $A=S_1(j_2)= \{ j_1 |  \mathbb{E}_{v_1,v_2}[(j_2v_2 + d_2)\cdot 1_{j_1v_1\geq j_2v_2}] 
\leq b_1,j_1 \in [0,\infty)\}$, $\forall j_2 \in [0,\infty)$ is non-empty, convex, compact. \end{lem}

\begin{proof}$S_1(j_2)$ is a strategy set and naturally non-empty. The product of all players' strategy sets are therefore also non-empty. For any given $j_2$, any combination of a feasible strategy's parameter still creates a feasible strategy (due to its linear form). Therefore $S_1(j_2)$ is convex. The set $S_1(j_2)$ contains all of its limits, therefore it is a closed set. Due to bidding range and budget, it is also bounded. The product of all players' strategy sets are therefore closed and bounded. According to Heine-Borel Theorem, the sets are compact.
\end{proof}

\begin{defi}: A set-valued function $u$ defined on a convex set $S_1(j_2)$ is quasiconcave if every upper level set of $u$ is convex, or $P_{j_1} = \{j_1\in S(j_2): u(j_1) \geq a\}$ is convex $\forall a \in \mathbb R$. \end{defi}

\begin{lem}\label{lem2.3} The correspondence $\varphi: S_1 \to 2^{S_1}$, where $\varphi(S_1)=\mathbf b_1$ is convex, $\forall s \in S_1$. \end{lem}

\begin{proof} First, we prove utility $u_i$ is quasiconcave. 

Let $\sigma_i^1, \sigma_i^2 \in \mathbf b_i$, since they are best responses, we have utilities $u_i^1=u_i(\sigma_i^1, \sigma_{-i}) \geq u_i(\tau_i,\sigma_{-i}), \forall \tau_i \in S_i$, and $u_i^2=u_i(\sigma_i^2, \sigma_{-i}) \geq u_i(\tau_i,\sigma_{-i}), \forall \tau_i \in S_i$. Hence, $\lambda u_i^1 + (1-\lambda)u_i^2 \geq u_i(\tau_i,\sigma_{-i}), \lambda \in [0,1]$. 

Given any $a \in \mathbb R$, if we create a upper level set $p_a$ containing all $j_1 \in S(j_2)$ that meet the condition of having a utility $u_i \geq a$, and if $p_a$ is always a convex set, then $u_i$ is quasiconcave. This is apparent, as $u_1(j_1)=E_3 -E_4=\mathbb{E}_{v_1,v_2}[(v_1-j_2v_2)\cdot 1_{j_1v_1\geq j_2v_2}]$ is continuous and non-decreasing in $j_1$. If $j_1v_1 \geq j_2v_2$ and $j'_1 v_1 \geq j_2v_2$, we would always have $\lambda j_1v_1 \geq \lambda j_2v_2$ and $(1-\lambda)j'_1 v_1 \geq (1-\lambda)j_2v_2$ for any $\lambda \in [0,1]$. Adding both sides of the inequation respectively: $(\lambda j_1 + (1-\lambda) j'_1) v_1 \geq j_2v_2$, which means $\lambda j_1 + (1-\lambda) j'_1$ is also a member of $p_a$, or that any $p_a$ is convex. 

Since the utility function $u_1$ is defined on convex set $S_1$ and all of its upper level set is convex, the utility function is quasiconcave. Also, as $u_i$ is quasiconcave, we have $u_i(\lambda \sigma_i^1 + (1-\lambda) \sigma_i^2, \sigma_{-i}) \geq \lambda u_i^1 + (1-\lambda)u_i^2 \geq u_i(\tau_i,\sigma_{-i})$. Therefore $\lambda \sigma_i^1 + (1-\lambda) \sigma_i^2$ is also a best response, it is in the $\mathbf b_i$ set. $\mathbf b_i$ is therefore convex-valued. Finally, $\varphi$ is convex if and only if each $\mathbf b_i$ is convex. Any combination of best responses will still be a best response.
\end{proof}

\begin{defi}[Upper hemicontinuity \cite{ok2007real}]\label{upperhemi} Correspondence $S: \Psi \to \Xi$ is upper hemicontinuous, if for every $\psi_1 \in \Psi$ and $\epsilon>0$, $\exists \delta>0$ s.t.: if $\psi_2 \in \Psi$ and $||\psi_2-\psi_1||<\delta$, then $S(\psi_2) \subset B_\epsilon(S(\psi_1))$, where $B_\epsilon(x)$ denotes the $\epsilon$-ball around $x$. Correspondence $S$ is lower hemicontinuous, if for any open set $U \subset \Xi$ with $S(\psi_1) \cap U \neq \emptyset$, $\exists \epsilon>0$, s.t. $\forall \psi_2 \in B_\epsilon(\psi_1)$, $S(\psi_2)\cap U \neq \emptyset$.\end{defi}

\begin{lem}\label{lem:s1_upperhemi} let bidder 2's feasible strategies $j_2$ be in a set $\Psi$, let bidder 1's strategies $A=S_1(j_2),j_2 \in \Psi$ be in a set $\Xi$. The correspondence: $S_1: \Psi \to \Xi$ is continuous at all $j_2$.
\end{lem}

\begin{proof}
$\forall j_2 \in \Psi$, and a $\epsilon$-ball around $S_1(j_2)$, we can find a range $\delta$ around $j_2$, s.t. any $j'_2 \in \Psi, ||j'_2-j_2||< \delta$, has $S_1(j'_2)$ within the $\epsilon$-ball around $S_1(j_2)$. This is apparent, since for any given best response parameter $j'_2$ in the neighborhood of $j_2$, the corresponding strategy set in $S_1(j_2)$ would be a set of $j'_1$ that is in the neighborhood of $j_1$ (upper hemicontinuous). It is proven in \cite{dutta1989maximum} that if the graph $G(S_1)$ is convex when $S_1(j_2)$ is monotone increasing, then $S_1$ is lower hemicontinuous. In our case, due to the linear form, and according to Lemma \ref{lem1.1}, $S_1$ is lower hemicontinuous. Therefore, $S_1$ is continuous \cite{dutta1989maximum}. 
\end{proof}

\begin{thm}[Berge's maximum theorem \cite{ok2007real}]\label{berge} Let $\Xi,\Psi$ be topological spaces, $u_1:\Xi \times \Psi \to \mathbb R$ be a continuous function on the product space, and $S_1: \Psi \to \Xi$ be a compact-valued correspondence s.t. $S_1(j_2) \neq \emptyset$, $\forall j_2 \in \Psi$. Define $u_1^*(j_2)=\sup\{u_1(j_1,j_2):j_1 \in S_1(j_2)\}$,  $\sup$ being the maximum operator of $u$, and the set of maximizers $S_1^*:\Psi \to \Xi$ by:
$S_1^*(j_2)=\arg \sup\{ u_1(j_1,j_2):j_1 \in S_1(j_2)\} = \{j_1 \in S_1(j_2): u_1(j_1,j_2)=u_1^*(j_2)\}$. If $S_1$ is continuous (i.e., both upper and lower) at $j_2$, then $u_1^*$ is continuous and $S_1^*$ is upper hemicontinuous with nonempty and compact values. \end{thm}

\begin{lem}\label{lem2.2} Correspondence $\varphi: S_1 \to 2^{S_1}$, where $\varphi(S_1)=S_1^*=\mathbf b_1$, is upper hemicontinuous with non-empty and compact values, and has a closed graph. \end{lem}

\begin{proof}
According to \ref{berge}, since $S_1$ is continuous (Lemma \ref{lem:s1_upperhemi}), non-empty and compact (Lemma \ref{lem1.1}), the correspondence $\varphi$ is upper hemicontinuous with non-empty and compact values. It is apparent that best response set is a closed subset of the strategy set $S$ on all $s \in S$. Therefore $b_i$ is closed-valued. A closed-valued upper hemicontinuous correspondence has a closed graph.
\end{proof}

Lemmas \ref{lem1.1}, \ref{lem2.3} and \ref{lem2.2} apply to the strategy sets of all players. According to the lemmas, we can prove that our setup meets the conditions of Theorem \ref{kakutani}, therefore the game has NE.

\subsection{Pareto optimality}
\label{appendix:paretoOptimal}

Valuation of the service request is a linear function of the resource needed: $v_1=g_1 \omega_1 + k_1,v_2=g_2 \omega_2+k_2$, $g,k$ are constants, $\omega$ is amount of resource required. The allocation rule under NE is: $A^*_{v_1,v_2}=1 \text{, if } j_1 v_1 + d_1 \geq j_2 v_2 + d_2 \text{, otherwise } 2$. Form of the condition is from best response form in appendix Sec.~\ref{appendix:ne}. We also assume that both bidders have at least some access to the resources, as a form of fairness. We define the fairness constraint to be: $\mathbb{E}[\omega_1|_{A_{v1,v2}=1}] / \mathbb{E}[\omega_2|_{A_{v1,v2}=2}]=\gamma \in \mathbb R_{>0}$.

\begin{thm}
The allocation $A^*_{v_1,v_2}$ maximizes overall resource allocation $\omega_1+\omega_2$, subject to the fairness constraint, when the valuations are linear functions of resources. Or, the NE of the game achieves optimal resource allocation.
\end{thm}

\begin{proof}
Find the Lagrangian multiplier $\lambda^*$ that satisfies the fairness constraint with NE allocation $A^*_{v_1,v_2}$. Define $g,k$ as: $g_1 = (1+\lambda^*)/j_1 \text{ , } k_1 =-d_1/j_1$, and $g_2 = (1-\gamma \lambda^*)/j_2 \text{ , } k_2 =-d_2/j_2$. Then we can rewrite the allocation: $A^*_{\omega_1,\omega_2} = 1 \text{, if } \omega_1 (1+\lambda^*) \geq \omega_2(1-\gamma \lambda^*) \text{, otherwise } 2$. The rest of the proof is the same as in \cite{sun2006wireless}.
\end{proof}






\end{document}